\RequirePackage{fix-cm}
\documentclass[a4paper]{svjour3}                     % onecolumn (standard format)
\smartqed  % flush right qed marks, e.g. at end of proof
\usepackage{graphicx}
\usepackage{braket}
\usepackage{amsmath}
\usepackage{amssymb, amsfonts}
\usepackage{mathrsfs}
\usepackage{array}
\usepackage{verbatim}

\usepackage{tikz}
\usepackage{tikz-3dplot}
\usepackage{appendix}
\usepackage{url}
\usepackage[round]{natbib}
\usepackage{ragged2e}

\usepackage{mathptmx}      % use Times fonts if available on your TeX system
\begin{document}

%\title{Quantum simulation of the ground--state Stark effect in small  molecules}
\title{Calculation of the ground–state Stark effect in small molecules using the variational quantum eigensolver}
%\subtitle{A case study using IBM Q}


\author{$ ^*$Carlos Tavares$^{1,2}$ \and
        Sofia Oliveira$^{2,3}$ \and 
        Vitor Fernandes$^{2,3}$ \and 
        Andrei Postnikov$^{4}$ \and 
        Mikhail I. Vasilevskiy$^{3, 5, 6}$%etc.
}

\authorrunning{Tavares, Oliveira, Fernandes, Postnikov and Vasilevskiy} % if too long for running head

\institute{Carlos Tavares\\
    ctavares@inesctec.pt\\
    \\
    $^1$High-Assurance Software Laboratory/INESC TEC, same address as $^2$ %\\
    \and \\ 
    $^2$Department of Informatics, University of Minho, Campus de Gualtar, Braga, Portugal %\\ 
    \and \\
    $^3$Department of Physics, University of Minho, Campus de Gualtar, Braga, Portugal %\\
    \and \\
    $^{4}$Université de Lorraine - LCP-A2MC, Metz, France %\\
    \and \\
    $^{5}$Centro de F\'{\i}sica, Universidade do Minho, Campus de Gualtar, Braga
4710-057, Portugal % \\
    \and \\
    $^{6}$International Iberian Nanotechnology Laboratory, Braga, Portugal
    }

\date{}

\maketitle

\begin{abstract}
{\justify As quantum computing approaches its first commercial implementations, quantum simulation emerges as a potentially 
ground-breaking technology for several domains, including Biology and Chemistry. However, taking advantage of quantum algorithms
in Quantum Chemistry raises a number of theoretical and practical challenges at different levels, from the conception
to its actual execution. 
We go through such challenges in a case study of a quantum simulation for the hydrogen (H$_2$) and lithium hydride (LiH)
molecules, at an actual commercially available quantum computer, the {\tt IBM Q}. The former molecule has always been a playground for testing approximate calculation methods in Quantum Chemistry, while the latter is just a little bit more complex, lacking the mirror symmetry of the former. Using the Variational Quantum Eigensolver (VQE) method,  we study the molecule's ground state energy versus interatomic distance, under the action of stationary electric fields (Stark effect). Additionally, we review the necessary calculations of the matrix elements of the \emph{second quantization} Hamiltonian encompassing the extra terms concerning the action of electric fields, using STO-LG type atomic orbitals to build the minimal {basis} sets.}


\end{abstract}

\section {Introduction}
The beginning of the twentieth century witnessed a revolution in Physics, which led to the development of Quantum Mechanics that proved the ability to solve problems of the Classical Physics at very small scales, and to predict accurately and elegantly the behaviour of sub-atomic particles. From the beginning, Chemistry has been a natural field of application for the  Quantum Mechanics, as quantum effects are relevant at molecular scale in many phenomena, originating the new field of Quantum Chemistry, -- see, e.g.,
\citet{levine2014quantum}. The same happens in Biology, where it is known that
quantum effects are relevant in several processes, and it is even believed they can help explaining several macro-phenomena in the life sciences \citep {abbott2008quantum}.  

However, looking through Quantum Mechanics to these disciplines faces major obstacles, as calculations rapidly become
in\-trac\-table with the size of the molecular systems  involved, even with the help of the most advanced classical computational tools. The concept of quantum simulation, idealized by \citet {feynman1982simulating} in the 80s and later refined by \citet {lloyd1996universal}, has raised expectations on the mitigation on some of these problems via achieving an exponential gain in simulation on quantum systems, with potential impact throughout all areas of Physics \citep {georgescu2014quantum}, including Quantum Chemistry \citep{Cao-Review2019} and the Life Sciences \citep {Bi-Xue_Wang2018}. Recently, as the ``second quantum revolution''\footnote{Technological revolution, in which \emph {ideal} quantum effects have a crutial role, with application in many areas, from health, to communication and information technology -- see \citet{Nielsen-Chuang-book} and \citet{Schumacher-book}.}
is coming of age, the first quantum computers are starting to emerge and become available to broad researcher's community, giving means to the fulfillment of the Feynman's vision.
Compared with classical computers, quantum devices are \emph {ultimately} expected
to perform Quantum Chemistry calculations more quickly and accurately, handling larger molecules than it is possible with classical algorithms. This ``quantum speedup'' may lead to the design and discovery of new pharmaceuticals,
materials, and industrial catalysts \citep{Sukin2018}. A number of successful cases are described in literature on the {efficient} calculation of properties of interest for Chemistry, such as the electronic structure of molecules, phase diagrams, or reaction rates \citep {lidar1999calculating, paesani2017experimental, aspuru2005simulated, lanyon2010towards}. Good reviews on the subject are available in \citet{Cao-Review2019}, or \citet {mcardle2020quantum}, the latter also involving the simulation of \emph {Hydrogen} and \emph {Lithium-Hydride} molecules.

The conceptualization of a quantum simulation, from theory to experiment, poses many challenges \citep {whitfield2011simulation}, with no general recipe to tackle them. We hope to contribute to the progress in this area by exploring the simulations of two molecular systems, hydrogen (H$_2$) and lithium hydride (LiH) on a commercially available quantum computer, the {\tt IBM Q}, accessed through the QuantaLab UMinho Academic Q Hub, and programmed using the {\tt QISKit} platform \citep {cross2018ibm}. The hydrogen molecule, the simplest existing one and also very important in nature, has been the natural test case of experimental and theoretical research. In particular, its ground state properties and the dissociation curve have recently been recalculated using advanced classical \citep{Vuckovic2015} and quantum \citep {Colless2018} algorithms (the latter with extension to excited states). In a recent work, \citet {arute2020hartree} describe Hartree-Fock calculations (done on the {\tt Google Sycamore} quantum processor) for linear chains of up to twelve hydrogen atoms and discuss resulting errors in the system’s energy, along with possible ways to mitigate these errors. Similar works are likely to appear now in rapidly growing numbers; their importance is not in an
increased speed or accuracy in tackling the corresponding quantum-chemical problems, as compared with established “conventional” algorithms, but the demonstration that these problems enter into the circle of practical feasibility for quantum computer. By the path of getting necessary experience in obtaining accurate and stable results for benchmark systems, testing different algorithms, the power of working quantum computers being simultaneously on the rise, the question of “quantum supremacy” may soon enough be posed while confronting problems of genuine challenge for contemporary quantum chemistry.

In this work, we extend the study of the H$_2$ molecule as a standard benchmark towards the case of asymmetric LiH, whose ground-state calculation requires the inclusion of p-type atomic orbitals. Moreover we investigate the steady-state electronic Stark effect, i.e. the ground state
energy shift in response to a stationary external electric field \citep {Gurav2018}. We try to elucidate the essence of the quantum simulation algorithms to the broad community of physicists and chemists who may find the original works on quantum computation too technical to follow. We
start from the definition of the molecular Hamiltonian, followed by its preparation for quantum simulation to the application of the Variational Quantum Eigensolver (VQE) method, as well as its implementation and testing on the IBM Q. 

The article is organized as follows: in Section \ref {basic_hamiltonians} we briefly introduce the Quantum Hamiltonian formalism for many-body systems, the Hartree-Fock approximation and the second quantization representation; in Section \ref{quantum computation} we explain the mapping onto a system of qubits and designing the quantum circuit corresponding to the initial Hamiltonian, and the working principle of the VQE. Section \ref {case_study} is dedicated to the case study of  H$_2$ and LiH molecules where we present and discuss the procedure details and results of the calculation of the dissociation curves in the presence of electric field. The last Section 
offers a summary and concluding remarks. The Appendix \ref {appendix_A} contains details of the necessary matrix element's calculation for this molecular setting, which is not commonly available in the literature.

\section{Quantum Chemistry background}
\label {basic_hamiltonians}
\subsection {Quantum Hamiltonian formalism}
In this section, we outline the basic principles 
of the formulation of molecular Hamiltonians and 
the latter's
``preparation'' for numerical calculation of electronic characteristics relevant for Physics
and Chemistry. This is the domain, albeit represented by a quite simplistic case,
of traditional Quantum Chemistry.
A good introduction to the subject has been offered, for instance, by \citet{levine2014quantum}
and \citet{szabo2012modern}. Here we briefly describe 
just a few concepts and approximations essential for the formulation of the computational problem to be solved using quantum tools. 

The Quantum Hamiltonian formalism, in the Schr\"o\-din\-ger's formulation, is centred at the Hamiltonian operator, 
$H = T + V$, $T$ being the kinetic energy of the constituent particles and $V$ the potential energy 
of all interactions and fields
in the system, both internal and external. The action of this operator on the system's wavefunction (WF), 
$\vert {\Psi }\rangle$, describes the latter's evolution,
\begin{equation}
\label{eq_shrodinger}
i\hbar \frac {\partial }{\partial t}\vert {\Psi (t)}\rangle = H \vert {\Psi (t)}\rangle \; ,
\end{equation}
or yields the total energy of the system if it is in a stationary state,
\begin{equation}
\label{eq_shrodinger_st}
H\vert {\Psi}\rangle  = E \vert {\Psi} \rangle  \;.
\end{equation}
The wavefunction $\vert {\Psi}\rangle$, beyond time, depends on other arguments (such as spatial coordinates and spin components) according to the \emph {representation} used. Usually there are several possible solutions to the equation, which correspond
to different values of the energy (energy levels or eigenvalues, $E_n$), which are 
discrete for a confined (or bound) physical system. These states, called 
stationary states or {\it eigenstates}, are denoted $\vert {\Psi }_n\rangle$, with the index $n=1,\dots,m$, in general, corresponding to
a set of so-called \emph{quantum numbers} that distinguish the eigenstates.
The set of eigenstates constitutes the eigenbasis of the system that can be seen
as a set of mutually orthogonal vectors in a Hilbert space of dimension $m$.
The quantum system is also allowed to be in a {\it superposition state},
\begin{equation}
\vert {\Psi} \rangle = \lambda_1 \vert {\Psi_1} \rangle + \lambda_2 \vert {\Psi_2} \rangle 
            + \hdots + \lambda_m \vert {\Psi_m} \rangle \;,
\label{eq:Psi-lambda1}
\end{equation}
whose energy is not well-defined (and, therefore, such a state is non-stationary).
According to the statistical interpretation of Quantum Mechanics originally proposed by M.~Born \citep {saunders2010many},
a \emph{measurement} of such a quantum state can \emph{randomly} yield one of the eigenvalues
of its energy, $E_n$, with the probabilities given by the squared amplitudes of the basis \emph {eigenstates} participating, $\vert \lambda_n \vert ^2$.

\subsection {Many-particle systems}\label{manybody_section}

The Schr{\"o}dinger equation for a system of non-interacting particles can be decomposed
into a set of uncoupled equations for each particle and the system's WF can be factorized. A combination of two non-interacting and  non-en\-tang\-led systems
can be described by applying the tensor product
on the two vector spaces,\footnote{
For interacting or entangled systems, the total WF cannot be written as a product of those
of its parts. Entangled parts of a system, even if they do not interact physically,
may not be described by a wave function, they only can be represented 
by a density matrix. Entanglement is out of scope of this article, 
the interested reader may refer to an appropriate textbook, e.g., that of ~\citet{Schumacher-book}.
}
with resultant basis given as follows:
\begin{align}
\label{eq_basis}
\vert {{\Psi}^{(1)}}\rangle \otimes \vert{{\Psi}^{(2)}} \rangle 
&= \sum^{M_1}_{\alpha} \sum^{M_2}_{\beta} \lambda _{\alpha}\mu _{\beta} 
{\vert {\Psi^{(1)}_{\alpha}} \rangle \otimes \vert {\Psi^{(2)}_{\beta}}}\rangle 
\nonumber \\
&= \sum^{M_1}_{\alpha} \sum^{M_2}_{\beta} { \lambda _{\alpha}\mu _{\beta} \vert 
{\Psi^{(1)}_{\alpha}\Psi^{(2)}_{\beta}}}\rangle \;.
\end{align}
In Eq. \eqref{eq_basis}, $\Psi_{\alpha}^{(s)}$ denotes an {eigenfunction} 
of a state ${\alpha}=1,\dots,M_s$ of the system $\Psi^{(s)}$ ($s=1,2$). 
The dimension of the product vector is ${\rm dim}(\Psi^{(1)}){\,\ast\,}{\rm dim}(\Psi^{(2)})=M_1\cdot M_2$.

When the particles constituting the system are \emph{identical}, their \emph{spin} becomes highly relevant. 
The spin, which is an {intrinsic} angular momentum of the particle, distinguishes two different
types of particles, bosons (e.g. photons) and fermions (e.g. electrons and protons). 
For fermions, the {\it Pauli exclusion principle} states that the system's WF must be antisymmetric
with respect to permutation of any two particles. It implies important restriction
upon the WF, namely that
the product vector \eqref {eq_basis}, if applied to a pair of non-interacting electrons,
is not compatible with the Pauli principle. 

In Quantum Chemistry, a single-electron WF is called orbital \citep{szabo2012modern}.
One can distinguish \emph {spatial orbitals} $\phi(\mathbf{r})$, where $r$ corresponds to spatial coordinates, and \emph {spin orbitals} $\chi(\mathbf{x})$,
where $\mathbf{x}=(\mathbf{r};s)$ and $s=\uparrow, \downarrow$ stands for two possible orientations of electron's spin. 
For two electrons, the Pauli principle means that 
\begin {equation}
\chi(\mathbf{x}_1,\mathbf{x}_2)=-\chi(\mathbf{x}_2,\mathbf{x}_1)
\end {equation}
or, equivalently,
\begin {equation}
\label {pauli_e}
\phi (\mathbf{r}_1, \mathbf{r}_2) =\mp \phi (\mathbf{r}_2, \mathbf{r}_1)\,,
\end {equation}
where the upper (lower) sign corresponds to parallel (antiparallel) spins of the two electrons.
If the electron-electron interaction is neglected, the correct (i.e. compatible with the Pauli principle)
two-electron WF is written in the form of the so-called \emph{Slater determinant},
\begin{eqnarray}
\vert {\chi^{(1)}_{\alpha}\,\chi^{(2)}_{\beta}}\rangle
=\frac 1 {\sqrt {2}} \left \vert \begin{array}{cc}
\chi _{\alpha} (\mathbf{x}_1) & \chi _{\beta} (\mathbf{x}_1)\\
\chi _{\alpha} (\mathbf{x}_2) & \chi _{\beta} (\mathbf{x}_2)
\end{array}\right \vert \;,
\label{eq:Slater_det}
\end{eqnarray}
where $\chi_{\alpha}(\mathbf{x})$ and $\chi_{\beta}(\mathbf{x})$ designate different spin orbitals.
A Slater determinant can be straightforwardly
generalized towards
the case of $N$ identical non-interacting particles.
It vanishes when any two electrons ``occupy'' the same spin orbital, as required by the Pauli exclusion principle.

The Slater determinant is a simple way of constructing a many-electron WF
from spin orbitals representing non-interacting electrons. Complete neglection of the Coulomb interaction
between the electrons would be too crude an approximation, while solving directly
the many-electron Schr{\"o}\-din\-ger equation is an intractable problem. A compromise is achieved
by a self-consistent field method also called {\it Hart\-ree-Fock} (HF) {\it approximation}. 
An effective one-electron operator is introduced, $v^{HF}(\mathbf{x})$, called {\it Fock operator},
which includes, as a part of the single electron potential energy, 
the electron's
interaction with all other electrons whose positions are averaged 
under an assumption 
that
the WF representing the system of $N$ electrons
is a single Slater determinant. An explicit expression for $v^{HF} (\mathbf{x})$ will be presented below. 

\subsection {Molecular Hamiltonian and Hartree-Fock approximation}
\label{subsec:MolHam_and_HF}

\begin{figure}[t!]
\centering
\medskip
 \includegraphics[width=.90\linewidth]{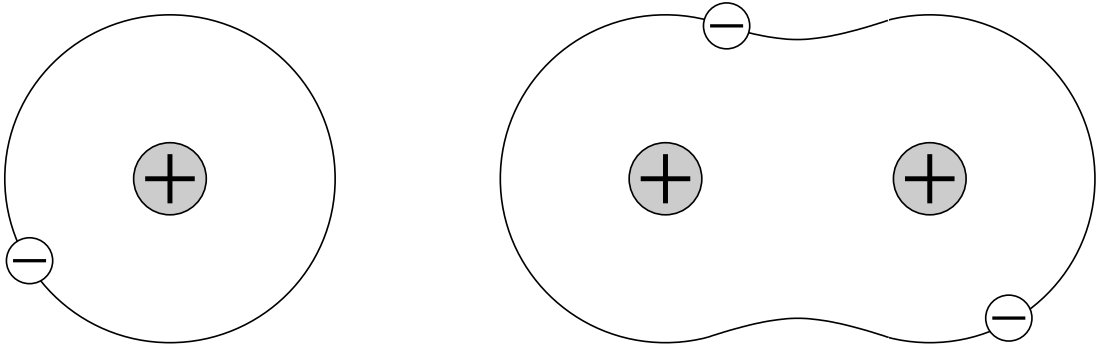}
\smallskip
\caption{Left: the hydrogen atom consists of a single electron and a proton
and has the energy of $-0.5$~a.u. in the ground state. Right: in the hydrogen molecule H$_2$, 
made of two nuclei and two electrons, the total energy can be lower than $-1$~a.u.,
which makes the molecule stable.}
\label{fig_h2_depiction}
\end{figure}

The general form of a molecular Hamiltonian is (in atomic units):
\begin{align}
\label{molecule_hamiltonian}
H_{\rm mol}
= -& \sum_{i=1}^N \frac {1}{2} \nabla_i^2 -\sum_{A=1}^M \frac {1}{2M_A} \nabla_{A}^2 
- \sum_{i=1}^N \sum_{A=1}^M \frac{Z_A}{r_{iA}} \nonumber \\
+& \sum_{i=1}^N \sum_{j>i}^N \frac {1}{r_{ij}}   
+\sum_{A=1}^M \sum_{B>A}^M \frac{Z_AZ_B}{r_{BA}}\,.    
\end{align}
The first and second terms of \eqref{molecule_hamiltonian} correspond to the kinetic energy
of the electrons (numbered by $i$ and $j=1,\dots,N$) and nuclei (numbered by $A=1,\dots,M$),
respectively. The third one represents the {Coulomb} attraction of each electron
to each nucleus with 
$r_{iA}$ being the electron-nucleus distance and $Z_A$ the nucleus charge.
Finally, the fourth and fifth terms correspond to the repulsion among the electrons and 
among the nuclei, respectively. It is common and well justified to use the {\it Born-Oppenheimer approximation}, 
which neglects the motion of the nuclei because they are much heavier than electrons,
whereby the potential energy of the nucleus-nucleus interactions 
becomes a constant  (for fixed placement of the nuclei) hence a parameter for the electron problem.
With this, the 
electron Hamiltonian \eqref{molecule_hamiltonian} reduces to:
\begin {equation}
\label{BO_hamiltonian}
H_{el} = - \sum_{i=1}^N \frac {1}{2} \nabla_i^2 - \sum_{i=1}^N \frac{Z_A}{r_{iA}}+ 
\sum_{i=1}^N \sum_{j>i}^N \frac {1}{r_{ij}}  \,.  
\end {equation}
For the H$_2$ molecule the Hamiltonian \eqref{BO_hamiltonian} depends
on a single parameter, the distance between 
the protons $d$. If the lowest eigenvalue of \eqref{BO_hamiltonian}, $E_0(d)<0$, is larger in absolute value
than the proton-proton repulsion energy, $E_{rep}(d)=d^{-1}$, the molecule is bound,
as illustrated in Fig.~\ref{fig_h2_depiction}.

%
% Figure 1
%

%\begin{figure*}[t!]
%\centering
%\begin{minipage}{.5\textwidth}
%  \centering
%  \includegraphics[width=.9\linewidth]{h.png}
%  \label{fig:test1}
%\end{minipage}%
%\begin{minipage}{.5\textwidth}
%  \centering
%  \includegraphics[width=.9\linewidth]{h2.png}
%  \label{fig:test2}
%\end{minipage}
%\caption{The hydrogen atom has a single electron moving around the proton
%and possessing the energy of $-0.5$~a.u. in the ground state. In hydrogen molecule, $H_2$, 
%there are two nuclei and two electrons moving around, and the total energy can be lower than $-1$~a.u.,
%which makes the molecule stable.}
%\label {fig_h2_depiction}
%\end{figure*}

The Hamiltonian \eqref{BO_hamiltonian} has to be reduced to a single-elec\-tron one in order to proceed
with finding its eigenvalues, which is achieved by means of the HF
approximation, where one 
takes an average over the positions and spins of all electrons but one (to be labelled by $i=1$).
This is done by multiplying \eqref{BO_hamiltonian} by 
$\vert {\chi^{(1)}_{\alpha}\chi^{(2)}_{\beta}\dots \chi^{(N)}_{\gamma}}\rangle$ 
and the corresponding ``bra'', both in the form of Slater determinants of dimension $N$ 
(the number of electrons in the system), and integrating over 
$\mathbf{x}_2,\, \mathbf{x}_3,\,\dots,\,\mathbf{x}_N$, which leads to :
\begin{equation}
\label{HF_equation}
\left (- \frac {1}{2} \nabla_1^2 - \sum_{A=1}^M \frac{Z_A}{r_{1A}}+ v^{HF}_{1}\right ) \chi_{\alpha } (\mathbf{x}_1) =\epsilon _{\alpha } \chi_{\alpha }  (\mathbf{x}_1)\,,  
\end{equation}
%https://www.overleaf.com/2973819279mfjwnddhczvh
where $v^{HF}_{1}$ is the average potential experienced by the ``chosen'' electron,
and $\epsilon_{\alpha}$ is the single-electron energy. The HF potential can be written
in the form:
\begin{eqnarray}
\label{HF_potential}
%\begin{aligned}https://www.overleaf.com/2973819279mfjwnddhczvh
v^{HF}_{1}&=&\sum_{\beta }\int\! \vert \chi_\beta  (\mathbf x_2)\vert ^2\frac{1}{|r_{12}|}\, d\mathbf x_2
\nonumber \\ 
%&&-\frac{1}{\vert \chi_\alpha  (\mathbf x_1)\vert ^2}
&&-\;\frac{\displaystyle
\sum_{\beta }
\int\! \chi_\alpha^{\ast} (\mathbf x_1)\chi_{\beta}^{\ast} (\mathbf x_2)
\frac{1}{|r_{12}|}\chi_\beta (\mathbf x_1)\chi_\alpha (\mathbf x_2) \, d\mathbf x_2}
{\displaystyle \vert \chi_\alpha  (\mathbf x_1)\vert ^2}\;.
%\end{aligned}
\label{HF_potential}
\end{eqnarray}
The two terms in Eq.~\eqref{HF_potential} are called Coulomb and exchange energies, respectively. The latter poses the main difficulty 
in solving Eq. \eqref{HF_equation};
however, its neglection (known as the Hartree approximation)
results in unsustainable error. Due to the nonlinearity of the HF approximation, the equations are solved 
in practice by self-consistent (iterative) methods, using a finite set of spatial basis functions, 
$\phi_\mu(\mathbf{r})$ ($\mu =1,2,$ $\dots$, $K$) -- see, e.g., \citet {szabo2012modern}. The solution yields a set HF spin orbitals $\{\chi_\alpha \}$ with corresponding energies $\{\epsilon _\alpha \}$, $\alpha =1,2,\dots,2K$. 
It must be $2K\ge{N}$, the number of electrons in the system. 
The possibilities to place $N$ electrons over $2K$ spin orbitals gives rise to $(2K)!/(N!(2K-N)!)$ Slater determinants, one of which represents the ground state of the system and the others correspond to excited states. The HF approximation takes into account the Quantum Mechanical correlation caused by the Pauli principle, however, only of electrons with parallel spins. The difference between the approximate HF energy and the exact energy of the system is known as \emph{correlation} correction (or energy).

It is common to use, as initial approximation
basis sets to represent molecular orbitals (MO) in the HF equations, 
the
linear combinations of atomic orbitals (LCAO).
Since the exact atomic orbitals 
for a given many-electron atom
are difficult to construct, the
so-called \emph{Slater-type orbitals} (STO) are 
sometimes
used,
which are
inspired by the (exactly known) 
radial asymptotics of
spatial orbitals of the 
hydrogen atom,\footnote{%
The STO include a simple power function of radius instead of a polynomial,
and hence do not possess radial nodes.}
$$\phi (\mathbf{r}) {\sim}r^{n-1}e^{-{\zeta}r}Y_{l,m}(\theta, \varphi)$$ (here $Y_{l,m}$ is a spherical harmonic). 
For instance,
one can use
$$
\phi^{\rm STO}_{1s}(\zeta,\mathbf{r}-\mathbf{R}_{\rm A})=
\left(\!\frac{\zeta^3}{\pi}\!\right)^{\!\!\tfrac{1}{2}}\!\!e^{-\zeta|\mathbf{r}-\mathbf{R}_{\rm A}|}
$$
for $s$-states, where $\zeta$ is the Slater orbital exponent. 
As the STO functions are difficult to handle in many-center integrals, one practical resort consists of approximating these functions with linear combinations of \emph{Gaussian functions}, known as STO-LG functions. The calculation of 
necessary matrix elements is then
greatly facilitated,
because the multi-center
integrals with Gaussian functions can be evaluated analytically (see Appendix \ref {appendix_A}).
In this work, a set of such functions with $n=3$ Gaussians mimicking each STO function, named STO-3G basis,
is used. For the 1$s$ state, such a function is: 
\begin{eqnarray}
&&\phi^{\rm STO-3G}_{1s}(\zeta,\mathbf{r}) =
c_1\!\left(\!\frac{2\alpha_1}{\pi}\!\right)^{\!\!\frac{3}{4}}\!\!e^{-\alpha_1 r^2} \nonumber \\
&& \quad\quad
+\; c_2\!\left(\!\frac{2\alpha_2}{\pi}\!\right)^{\!\!\frac{3}{4}}\!\!e^{-\alpha_2 r^2}   +\; c_3\!\left(\!\frac{2\alpha_3}{\pi}\!\right)^{\!\!\frac{3}{4}}\!\!e^{-\alpha_3 r^2}\,. 
\label{sto3g}
\end{eqnarray}
Here $\alpha_i$ are the Gaussian orbital exponents that have been optimized
for the best possible  approximation of $\phi^{\rm STO}_{1s}(\zeta,\mathbf{r})$ for a given $\zeta$ \citep {Hehre1969}.
The corresponding spin orbitals, $\chi_\alpha(x)$, are obtained from $\phi^{\rm STO-3G}_{\mu}$ by multiplying them
with a spinor $\psi(s)$,  $s =\uparrow ,\, \downarrow $.

\subsection {Second quantization}
In the quantum mechanics of systems consisting of a number of identical particles (electrons, in our case),
it is common to use the formalism called \emph{second quantization}, originally introduced
by P.~Dirac -- see, e.g., \citet{dirac1981principles}. This formalism deals with the whole system of particles, instead of each particle individually, by introducing a new way of describing states, 
by the latter's \emph{occupation} \emph{numbers}.
Let $\{\chi_\alpha(\mathbf x)\}$ be a complete set of one-electron (atomic or molecular) spin orbitals that constitute
the Hilbert space of a single particle. If the particles were non-interacting bosons, a state of the whole system
could be entirely specified by indicating the numbers of particles, $n_\alpha$, occupying each of these orbitals. Such
an occupation number state can be designated by a state vector $\vert{n_1,n_2,...} \rangle$.
If the particles interact with an external field or with each other (but still assuming that they are bosons
and no restrictions are imposed by particle's spin), the state vector in the occupation number representation 
will evolve with time, obeying the time-dependent Schr\"odinger equation \eqref{eq_shrodinger} with the Hamiltonian
written in the occupation numbers representation:
\begin{equation}
\label{2q-H} 
H=H_1+H_2= \sum_{\alpha,\beta} \tau_{\alpha\beta} a_{\alpha}^{\dagger} a_{\beta}+\frac{1}{2} \!
\sum_{\alpha,\beta,\atop\gamma,\delta} \mu_{{\alpha}{\beta}{\gamma}{\delta}} a_{\alpha}^{\dagger} a_{\gamma}^{\dagger} a_{\delta} a_{\beta}\,.
\end{equation}
The summation is over states in the single-particle Hilbert space, e.g., $1s$-, $2p$-like, etc.,
$\tau_{\alpha\beta}$ being a matrix element of the single-electron energy, 
\begin{equation}
\label{one_int} 
\tau_{\alpha\beta}=\int d\mathbf x_1 \chi_\alpha^* (\mathbf x_1)\left(\frac{-\nabla^2}{2}+\sum_A  \frac{Z_A}{|r_{A1} |}\right) \chi_\beta (\mathbf x_1)\,.
\end{equation}
The second term in \eqref {2q-H} represents the  Coulomb interactions between the particles,
with the matrix element given [according to the convention used in Quantum Chemistry \citep{szabo2012modern}] by:
\begin{equation}
\label{two_int}
\mu_{{\alpha}{\beta}{\gamma}{\delta}}=\int d\mathbf x_1 d\mathbf x_2 \chi_\alpha^* (\mathbf x_1)\chi_\beta (\mathbf x_1)\left(\frac{1}{| r_{12}|}\right) \chi_{\gamma}^{*} (\mathbf x_2)\chi_\delta (\mathbf x_2)\,.
\end{equation}
The integration in Eqs. \eqref{one_int} and \eqref{two_int} is over coordinates (and summation over spins) 
of one or two electrons labelled 1, 2. 

The Hamiltonian \eqref {2q-H} is written in terms of so-called  
\emph{creation}, $a^{\dagger}$, and \emph{annihilation},  $a$, operators, which add one particle
to (or, remove from) an orbital $\alpha$, respectively:
\begin{eqnarray}
\label{ac_operators}
\nonumber
%\begin{aligned}
a_{\alpha}^{\dagger}\,\vert {n_1,n_2,\hdots}\rangle &=& \sqrt {n_\alpha\!+\!1}\, \vert n_1,n_2,{\hdots}\rangle\, ;\\ 
a_{\alpha}          \,\vert {n_1,n_2,\hdots}\rangle &=& \sqrt {n_\alpha}\,       \vert n_1,n_2,{\hdots}\rangle\, .
%\end{aligned}
\end{eqnarray}
The product $a_{\alpha}^{\dagger}a_\alpha$ is the occupation number operator for the orbital $\alpha$. 
In the case of bosons, the creation and annihilation operators for different $\alpha$  and $\beta$ commute,
because different orbitals are filled independently.  These is not the case for fermions,
because of the Pauli exclusion principle. By virtue of this, the following (anti-commutation) relations hold 
for the electron operators:
\begin{equation}
\label{anticom}
a_{\alpha}\,a_{\beta}^{\dagger} + a_{\alpha}^{\dagger}\,a_{\beta} = \delta_{{\alpha}{\beta}}\,.
\end{equation}
It can be shown that \eqref{anticom} guarantees that the occupation numbers can take only values 0 and 1
in accordance with the Pauli principle \citep{dirac1981principles}.
Therefore, the Hamiltonian \eqref{2q-H} has the same form for bosons and fermions, 
the only difference being
in the (anti-)commutation relations of the creation and annihilation operators. For fermions, each state
$\vert{n_1,n_2,...} \rangle$ of this Hamiltonian corresponds to a Slater determinant
in the \emph{Fock space} (of dimension $2K$), with the number of columns and rows equal to the number of electrons
in the system, $N=\sum_{\alpha=1}^{2K} n_\alpha$. 

The choice of single-electron basis functions $\chi_\alpha^{\ast}(\mathbf x)$ is, in principle, arbitrary,
but if we ``guess'' their form close to the ``true'' WFs of the system 
(which actually are not well-defined in the single-electron form!), the non-diagonal elements
of the matrices $\tau_{\alpha\beta}$ and $\mu_{{\alpha}{\beta}{\gamma}{\delta}}$ will be much smaller
than the diagonal ones. 
For practical calculations of these integrals, the basis functions are expressed in 
terms of the STO-3G 
sets explained in the previous section. 
The choice of molecular orbitals is based on the MO-LCAO approximation. One can improve this initial approximation 
by solving first the HF equation \eqref{HF_equation} and using its solutions to calculate 
the matrix elements. Then the diagonalization of Eq. \eqref{2q-H} amounts to the evaluation of the correlation energy.    

In this article we are going to consider also  the stationary Stark effect described by the following 
(single-electron) Hamiltonian:
\begin{equation}
\label{Stark-Ham}
H_{S} = -\mathbf{E}\cdot \mathbf{r}\,,
\end{equation}
where $\mathbf{E}$ is the electric field intensity.
Its second-quantization representation is identical to $H_1$ in \eqref{2q-H}, and the corresponding matrix element is written as 
\begin {equation}
\label {Stark_int} 
\tau_{\alpha\beta}^{S}=\int d\mathbf x_1 \chi_\alpha^{\ast}(\mathbf x_1)\left(-e\,\mathbb{E}\,r\cos{\theta}\right)\chi_\beta (\mathbf x_1)\,,
\end {equation}
where $\mathbb{E}=\vert \mathbf{E}\vert $ and $z$-axis is assumed to be directed along $\mathbf{E}$.
The use of second quantization formalism is facilitated, for instance, by the {\tt PyQuante} \citep {muller2017pyquante} 
and the {\tt PyScf} \citep {sun2018pyscf} tools, Python libraries targeted to quantum chemistry calculations.
We present the matrix elements \eqref{one_int}, \eqref{two_int} and \eqref{Stark_int} calculated for $1s$, $2s$ and $2p_z$ atomic orbitals 
in the Appendix \ref {appendix_A}.

\section{Quantum simulation of a Quantum Chemistry Hamiltonian}
\label{quantum computation}

\subsection{Mapping the fermion Hamiltonian onto a qubit representation}
\label {qubit_representation}
%
% Figure 2
%
\begin{figure}[!b]
\centering
\includegraphics[width=0.90\linewidth]{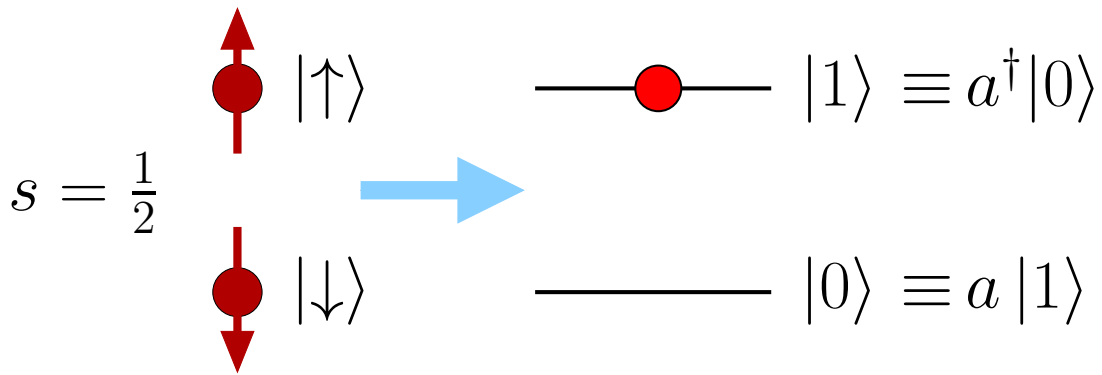}
\smallskip
\caption {A scheme illustrating the mapping of a fermion onto a qubit. The arrows indicate two spin projections. The two states of the Hamiltonian $\hat {H} = C\hat {\sigma}_z$ ($C$ is a \emph {positive} real constant) are usually denoted as $\ket {0}$ (lower energy state) and $\ket {1}$.They are connected by the operators $a$ and $a^{\dagger}$. Considered as spin states, they may be denoted as $\begin {pmatrix} 0 \\ 1 \end {pmatrix} = \ket {0}$ and $\begin {pmatrix} 1 \\ 0 \end {pmatrix} = \ket {1}$. It leads to the correspondence between the $a$ and $a^{\dagger}$ operators and the Pauli matrices given by Eqs. \eqref {J-W1} - \eqref {J-W} below.} 
\label{fig:J-W}
\end{figure}

In order to perform quantum computations, one needs to map the second-quantization Hamiltonian
onto a qubit (spin) representation and then design the corresponding quantum circuit that implements it.
The basic idea is to replace the fermionic operators $a$ and $a^{\dagger}$ with tensor products of the Pauli matrices,
\begin{equation*}
\sigma _x   = \left[\begin{array}{cc} 0 & 1 \\ 1 & 0 \end{array}\right]\,,\qquad  
\sigma _y   = \left[\begin{array}{cc} 0 & -i \\ i & 0 \end{array}\right]\,,\qquad
\sigma _z   = \left[\begin{array}{cc} 1 & 0 \\ 0 & -1 \end{array}\right]\,,
\end{equation*}
which can be done in a number of ways, such as the Jordan-Wigner or Bravyi-Kitaev transformations \citep{Cao-Review2019}.
The former, addressed in this section, is a specific method based on the isomorphism between the creation and annihilation operators and the algebra of the Pauli matrices \citep {whitfield2011simulation}.

In the case of a single (one-electron) state, the Jordan -- Wigner (JW) mapping is simple. Following the convention of Fig. \ref {fig:J-W}, common in Physics, 
\begin{eqnarray}
%\quad
&& a^{\dagger} {\;\Leftrightarrow\;} \sigma^+\,{\equiv}
\frac{1}{2}\left(\sigma_x +i \sigma_y \right) = \left[\begin{array}{cc} 0 & 1 \\ 0 & 0 \end{array}\right]\,; \label {J-W1}\\
&& a {\;\Leftrightarrow\;}  \sigma^-\,{\equiv}\frac{1}{2}\left(\sigma_x -i \sigma_y \right)=
\left[\begin{array}{cc} 0 & 0 \\ 1 & 0 \end{array}\right]\,; \\
%\quad
&& a^{\dagger}a-\tfrac{1}{2} {\;\Leftrightarrow\;} \frac{1}{2}\sigma_z    =
\left[\begin{array}{c@{\!\;\,}c} \tfrac{1}{2} & 0 \\*[2pt] \;0 & -\tfrac{1}{2} \end{array}\right]\,.
\label {J-W}
\end{eqnarray}
The matrices $\sigma^{\pm}$ represent
the spin-raising and spin-lowering operators, respectively, while $\sigma_z$ is  related to the occupation number operator. 

However, usually another convention is used in quantum information, as the computational basis is defined as follows:

\begin {equation*}
\ket {0} = \begin {pmatrix} 1 \\ 0 \end {pmatrix}\,;\qquad \ket {1} = \begin {pmatrix} 0 \\ 1 \end {pmatrix}\,.
\end {equation*}

\noindent Accordingly,
\begin {equation}
a^{\dagger} \Leftrightarrow \hat {\sigma}^{-} \qquad \text {and} \qquad a \Leftrightarrow \hat {\sigma}^{+}\,.
\label{a-sigma}
\end {equation}
In case of $N>\!\!1$ fermions, the mapping becomes slightly more complex. In order to satisfy the {anti-commutation relations} \eqref{anticom}
between \emph{any} pair of fermionic operators,  
one numerates the states by a single index ($\alpha$) and adds the \emph{string}, 
i.e. [spin]=[fermion]$\times${[string]}, taking into account the occupation numbers, $n_\alpha$, of states with $\beta < \alpha$, for a given $\alpha$:
\begin {equation}
\label {J-W-2}
%\quad
\sigma^+_\alpha {\;\Leftrightarrow \;} a_\alpha {e^{i\pi\sum_{\beta<\alpha}n_\beta}}\,, \qquad 
\sigma^-_\alpha {\;\Leftrightarrow \;} a^{\dagger}_\alpha {e^{-i\pi\sum_{\beta<\alpha}n_\beta}}\,.
\end {equation}
The relation \eqref {J-W-2} holds for multiple fermions and the phase factors (compare to (\ref{a-sigma})) can be represented by the Pauli matrices, $(\sigma _z)_\beta$, acting on the corresponding fermionic states. Therefore, the fermionic operators are mapped onto direct products of Pauli matrices as follows:
\begin{eqnarray}
%\quad
a_\alpha\, &{\Leftrightarrow}& \mathbf{1}^{{\otimes}(\alpha -1)}{\otimes}
\,(\!{\sigma^+}\!)_\alpha {\otimes}
\,\!(\!\sigma _z\!)^{{\otimes}(N-\alpha ) } \nonumber \\
&&=
\left[\!\!\begin{array}{c@{\!\;\;}c}1 & 0 \\ 0 & 1 \end{array}\!\!\right]^{\!{\otimes}(\alpha -1)}
\!\!\!{\otimes}
\left[\!\!\begin{array}{c@{\!\;\;}c}0 & {1} \\ 0 & 0 \end{array}\!\!\right]_\alpha
\!{\otimes}\!
\left[\!\!\begin{array}{cc}1 & 0 \\ 0 & \!\!\!-1 \end{array}\!\!\right]^{\!{\otimes}(N-\alpha )}\,; 
\\*[3pt]
a_\alpha^{\dagger}\, &{\Leftrightarrow}& \mathbf{1}^{{\otimes}(\alpha -1)}{\otimes}
\,(\!{\sigma^-}\!)_\alpha{\otimes}
\,\!(\!\sigma _z\!)^{{\otimes}(N-\alpha) } \nonumber \\
&&=
\left[\!\!\begin{array}{c@{\!\;\;}c}1 & 0 \\ 0 & 1 \end{array}\!\!\right]^{\!{\otimes}(\alpha -1)}
\!\!\!{\otimes}
\left[\!\!\begin{array}{c@{\!\;\;}c}0 & 0 \\ {1} & 0 \end{array}\!\!\right]_\alpha
\!{\otimes}\!
\left[\!\!\begin{array}{cc}1 & 0 \\ 0 & \!\!\!-1 \end{array}\!\!\right]^{\!{\otimes}(N-\alpha )}\,.
\label {J-W-3}
\end{eqnarray}  
Thus, any Hamiltonian operator written in the second quantization representation can be rewritten
in terms of the raising and lowering spin operators and the Pauli matrix $\sigma_z$. 
A catalogue of such translations can be found in Table A2 of the work 
by \citet{whitfield2011simulation}. For a Hilbert space of $2K$ spin orbitals, a system of $2K$ fermions (i.e. qubits) is required for the JW mapping. The resulting qubit Hamiltonian has the following generic form:
\begin {equation}\label {eq_H-generic}
 {H} = \sum_{i;\, q} h_{q}^{i} {{\sigma}_{i}^{(q)}} + \sum_{i_1,i_2;\, q_1,q_2} h_{q_1,q_2}^{i_1,i_2} {\sigma}_{i_1}^{(q_1)}\otimes {\sigma}_{i_2}^{(q_2)} + \cdots
\end {equation}
where the indices $i$ mean the type of the Pauli matrix ($x$, $y$ or $z$), the indices $q$ run over qubits and $h$ are some coefficients. This form is useful for the algorithms discussed in the next section.

\subsection {Quantum computation of the \emph {eigenvalues} of a Hamiltonian}
\label {sec-ground}

Once the molecule's Hamiltonian has been transformed into the qubit representation, the ground state energy can be evaluated using several methods. 
One of such methods 
where the \emph{quantum advantage} seems likely
is the calculation of \emph{eigenvalues} of Hamiltonians through the application of the 
\emph{quantum phase estimation} (QPE) algorithm \citep{luis1996optimum}, which also has 
several other applications, such as in the resolution of linear equations \citep {harrow2009quantum}.
The method requires an approximation of the evolution operator, $\hat U=\exp {(-i{H}t})$ ($t$ is time), and applying it to the initial state an appropriate number of times. {For an eigenstate, the application of $\hat U$ results in adding a phase $(-Et)$, so that the energy eigenvalue $E$ can be estimated.} 
Unfortunately, despite its theoretical attractivity
and a broad scope of possible applications, the method poses serious technical difficulties, which makes its practical realisation unlikely
at the present level of maturity of quantum computers. Namely, the QPE method requires a very large number of entangled qubits and quantum gates to be effective. 

Alternatively, one can adopt a strategy of applying the Hamiltonian over a state several times, measuring the result (i.e., performing the \emph {quantum sampling}), in order to obtain an estimation of the expected \emph{eigenvalue},
for which effective algorithms are available, 
particularly the \emph {Quantum Expected Eigenvalue Estimation} (QEE) method.
The method requires that the Hamiltonian operator can be decomposed into a \emph{polynomial} ($M$) independent $n$-qubit operators as exemplified by Eq. \eqref {eq_H-generic}
and consists in the ``measurement'' of the expectation values of 
such operators for a trial state $\vert \Psi \rangle$ (also known as the \emph{ansatz}):
\begin {eqnarray}
\nonumber
\braket {H} &=&\langle \Psi \vert H \vert \Psi \rangle \\
&=&\sum_{i;\, q} h_{q}^{i} \braket {{\sigma}_{i}^{(q)}} + \!\!\sum_{i_1,i_2;\atop q_1,q_2}\!\!
h_{q_1,q_2}^{i_1,i_2} \braket {\sigma_{i_1}^{(q_1)}\otimes \sigma_{i_2}^{(q_2)}} + \cdots
\label {eq_i}
\end {eqnarray}
The estimation of the expectation values, $\langle \cdots \rangle$, requires repeated measurements with a large number of qubits but, on the other hand, the computational effort amounts to the evaluation of a polynomial number of independent terms.

\begin {table}[t!]
\caption{
\label{compqeeqpe}
Comparison of resources needed for two methods, QPE and QEE.
$M$: the number of independent terms of the Hamiltonian approximation,
$p$: the precision chosen, $O(...)$: \emph {assymptotic} lower bound of the associated resource function.
See text for details.}
\smallskip
\begin{tabular}{llll}
\hline
\rule[-8pt]{0pt}{20pt}
\parbox[c]{0.12\linewidth}{~\\*[-1pt]Method} & \parbox[c]{0.25\linewidth}{Number of\\*[-1pt] state preparations} & 
\parbox[c]{0.16\linewidth}{Coherence\\*[-1pt] time} & \parbox[c]{0.19\linewidth}{Number\\*[-1pt] of steps} \\
\hline
\rule[0pt]{0pt}{11pt} QEE & $O(M)$ & $O(1)$      & $O({|h_{max}|}^2 M p^{-2})$ \\
\rule[-4pt]{0pt}{8pt} QPE & $O(1)$ & $O(p^{-1})$ & $O(p^{-1})$ \\
\hline
\end{tabular}
\end {table}

An objective comparison of the QPE and QEE methods 
is presented by \citet{mcclean2016theory} and summarized in Table~\ref{compqeeqpe}. One main advantage of the QEE, when compared with QPE, is that it largely reduces the need for gates,
but, more important, -- the amount of time the \emph {entanglement} over sets of qubits has to be maintained, i.e. the \emph {coherence time}, is $O(1)$ (\emph {independent of precision, $p$}), which is within grasp of existing quantum computers, while it grows \emph {linearly} with $p$, $O(p^{-1})$, for QPE. However, QEE introduces the need to prepare more copies of the  \emph{ansatz} to maintain the independence
of the terms in Eq. \eqref{eq_i} -- $O(M)$ against $O(1)$ for QPE, -- 
requiring polynomially more memory, i.e. more qubits.
Moreover, for a desired precision $p$, the number of 
necessary \emph {sampling} steps is $O({|h_{max}|}^2 M p^{-2})$, where $h_{max}$ is the term with the maximum norm in the decomposition of the Hamiltonian. In summary, the QEE method reduces the required minimum \emph{coherence} but introduces a polynomial complexity penalty, both in terms of memory and in terms of the number of steps necessary. Yet, it still holds an exponential advantage when compared to classical methods.

\subsection {Trial wave functions (\emph{ans\"atze})}\label {twavef}

The ground state energy estimation requires an appropriate \emph{ansatz}. 
If the number of electrons in the system, $N$, is fixed, one may
use the Slater determinant solution of the HF problem for the considered molecule, corresponding to its ground state. We shall denote it by $\vert{\Psi_0}\rangle$ and it may be written as 
\begin{equation}
\vert{\Psi_0}\rangle =\prod _\alpha ^{N} a^{\dagger}_\alpha \vert \text {vac}\rangle \,,
\nonumber
\end{equation}
where $\alpha$ runs over occupied orbitals and $\vert \text {vac}\rangle$ denotes vacuum (with no particles). 
Alternatively, one may 
start by defining a new ``vacuum'' state in the $N$-particle sector of the Fock space, which can be chosen as $\vert{\Psi_0}\rangle$ and used to prepare the parametrized trial quantum state  \citep{barkoutsos2018quantum}.
It can be done by a quantum circuit implementing a unitary operator, $\hat U$, that represents a set of perturbations to the state $\vert{\Psi_0}\rangle$: 
\begin{equation}
\vert \Psi(\overrightarrow{\theta})\rangle = \hat U(\overrightarrow{\theta}) \vert \Psi_0\rangle\, ,
\label{ansatz}
\end{equation}
The parametrized ansatz will be used to estimate the energy with respect to the Hamiltonian. Here $\overrightarrow {\theta}$ stands for
the whole set of parameters (also called ``gate angles'' in this context) that can be adjusted and used in the optimization procedure (see Sec. \ref {vqe_method} below).  

There are several possible choices of constructing this operator, leading e.g.
to the so-called Unitary Coupled Cluster (UCC) and Heuristic approaches that 
have been over\-viewed
by \citet{Cao-Review2019} and \citet{barkoutsos2018quantum}. There are options of choosing different 
\emph{ans\"atze} implemented in the {\tt QISKit} package.
Let us briefly consider the UCC approach, which has mainly been used in this work.

A flexible way to generate \emph{multideterminantal} 
(hence overcoming the HF approximation) reference states within the Coupled-cluster (CC)
method, suggested by \citet{jeziorski1981coupled},
has been translated by \citet{barkoutsos2018quantum}
(specifically under an angle of quantum algorithms for electronic structure calculations)
into the \emph{unitary version} of the CC approach (UCC). The operator {acting on the ``vacuum state'' according to Eq. \eqref{ansatz}} is chosen as follows:
\begin{equation}
\label {unitaryop}
\ket{\Psi(\overrightarrow{\theta})} = e^{\hat{T}(\overrightarrow{\theta})-\hat{T}^{\dagger}(\overrightarrow{\theta})}
\ket{\Psi_0}\,.
\end{equation}
Here $\hat{T}$ is an operator representing excitations from occupied to unoccupied states (labeled below by Greek and Latin indices, respectively), composed of hierarchical terms, 
\begin{equation}
\hat{T}=\hat{T}_1+\hat{T}_2+\dots \,,
\nonumber
\end{equation}
corresponding to $n$-particle excitations, namely, 
\begin{eqnarray}
\hat{T}_1 (\overrightarrow{\theta}) &=& \sum_{\alpha,a} \theta_{\alpha}^{a}{a}_a^{\dagger}\,{a}_\alpha\,,
\label{eq:T_1} \\
\hat{T}_2 (\overrightarrow{\theta}) &=& \frac 1 2 \sum_{\alpha,\beta;\;a,b} \theta_{\alpha\,\beta}^{a\,b}a_a^{\dagger}\,a_b^{\dagger}\,{a}_\alpha\,{a}_\beta\,,
\label{eq:T_2}\\
&\cdots & \nonumber
\end{eqnarray}
The UCC ansatz usually retains only the two first
terms in the expansion of $\hat{T}$, i.e. neglects 3-particle and higher order excitations. 
The expansion coefficients in \eqref{eq:T_1}, \eqref{eq:T_2} can be interpreted as matrix elements of a certain excitation operator between occupied and unoccupied orbitals. They can be assumed \emph {real}, i.e., $\{\theta_{\alpha}^{a},\; \theta_{\alpha\,\beta}^{a\,b}, \hdots\} \in \mathbb {R}$. 

The anti-Hermitian combination $\hat{T}-\hat{T}^{\dagger}$ in \eqref {unitaryop} makes the exponential operator unitary.
Unitary operations are natural on quantum computers, yet the implementation into quantum circuits is not that straightforward because of the non-commutation of different parts of the Hamiltonian, so the order in which the different terms are written in the exponent is important. This difficulty is bypassed by using the Trotter identity:   
\begin{equation}
\label {Trotter}
e^{(\hat{A}+\hat{B})}=\lim_{n{\rightarrow}\infty}\left[e^{\hat{A}/n}{\otimes}e^{\hat{B}/n}\right]^n\,,
\end{equation}
where $\hat{A}$ and $\hat{B}$ are two non-commuting operators, e.g.
$\hat{A}=\hat{T}_1-\hat{T}_1^{\dagger}$ and $\hat{B}=\hat{T}_2-\hat{T}_2^{\dagger}$.
Exact in the limit $n{\rightarrow}\infty$, it is an approximation for finite $n$.
Different Trotter approximations of the operator \eqref {unitaryop} can be implemented on a quantum computer by transforming it to the qubit representation and using standard circuit compilation techniques for the ``exponentiation'' of the Pauli matrices \citep {Cao-Review2019}. Some examples of such circuits and comparison of results obtained for different orders ($n$) of the Trotter approximation can be found in the work by \citet{barkoutsos2018quantum}.  

\subsection {Variational Quantum Eigensolver}
\label {vqe_method}

The variational method for the calculation of the ground state energy, also known in Physics as the Rayleigh-Ritz method, has widely been used for a long time in Quantum Chemistry -- see, e.g., \citet{levine2014quantum}. It is an approximation method used to estimate the lowest eigen\-value (the ground state energy) of a Hamiltonian,
\begin {equation}
\label {c_variational}
E [\Psi(\overrightarrow {\theta})]= \frac {\langle {\Psi(\overrightarrow {\theta})|H |\Psi(\overrightarrow {\theta})\rangle }}{\langle {\Psi(\overrightarrow {\theta})|\Psi(\overrightarrow {\theta})\rangle}}\, .
\end {equation}
The optimization consists in the determination of the set of parameters $\overrightarrow {\theta}$ that minimize the $E$ function.

In the hybrid quantum-classical algorithm implemented as the \emph {Variational Quantum Eigensolver} (VQE), the quantum computer prepares the parametrized trial function $\Psi(\overrightarrow {\theta})$, as discussed in section \ref {twavef}, and evaluates the energy with respect to the system's Hamiltonian, as discussed in section \ref {sec-ground}. Then a classically implemented algorithm updates the parameters $\overrightarrow {\theta } \in \mathbb {R}^n$ of the quantum state using a classical optimization routine,  and then repeats the previous step until convergence criteria (e.g., in energy and/or iteration number) are satisfied.{
{Any} optimization method capable of {performing this task}
can, {in principle,}
be used. On {\tt IBM Q} \citep{cross2018ibm}, a few methods for this purpose are available,
for instance, the \textit{Simultaneous Perturbation Stochastic Approximation Algorithm} %(SPSA)
\citep[SPSA, see][]{bhatnagar2012stochastic}, 
{caracterized by a}
very good performance under 
noise, or the {\tt Cobyla} method \citep{powell2007view}.} 

\begin{figure}[!t]
\smallskip
	\includegraphics[width=0.95\linewidth]{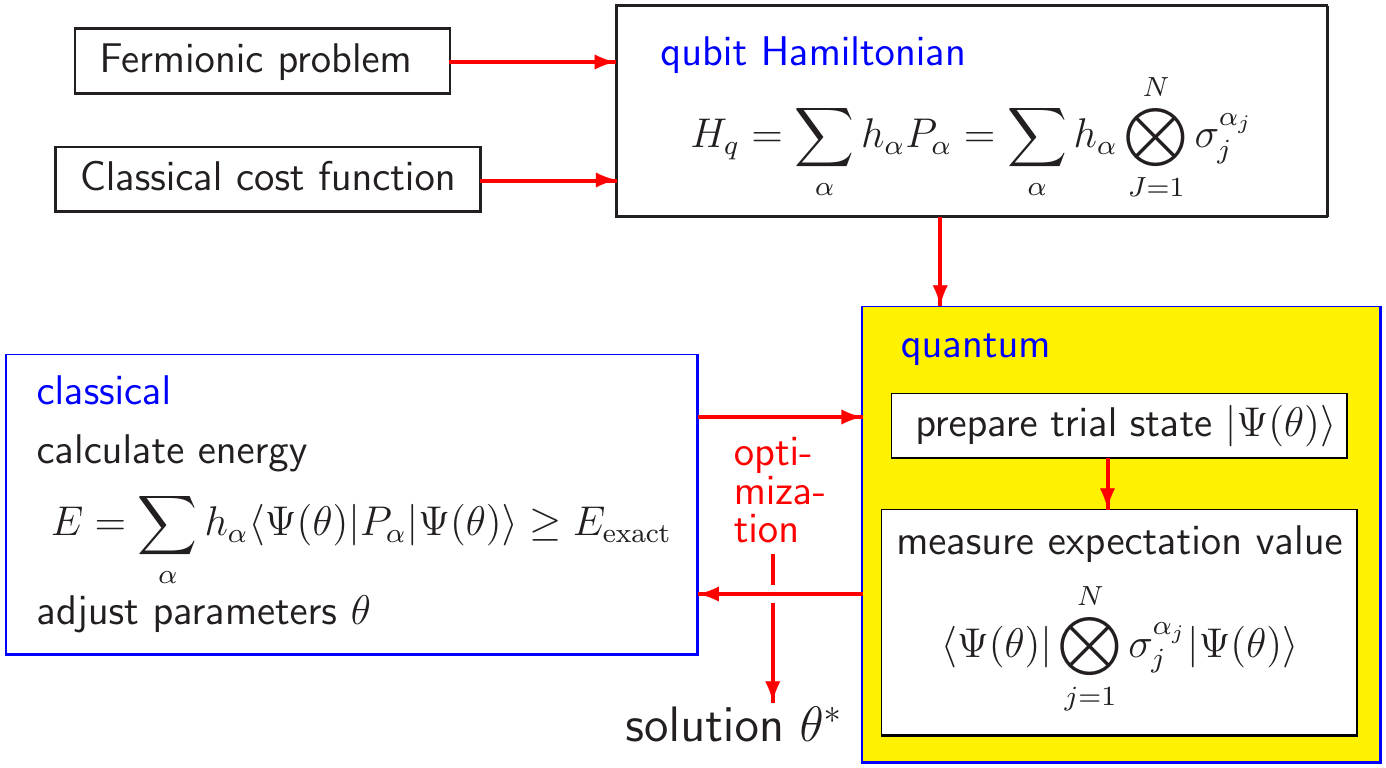}
	%\smallskip
	\caption{Application of the variational method to fermionic problems, adapted from \citet{moll2017quantum}.}
	\label{fig:variational_method}
\end{figure}

The VQE was introduced by \citet{peruzzo2014variational} {and applied since then in a number of quantum simulation / optimization
tasks -- see, e.g.,  ~\citet{moll2017quantum}. The scheme of the method is
depicted in Fig.~\ref {fig:variational_method}, adapted from the latter work.}
A good additional discussion of this method can be found in the work by \citet{mcclean2016theory}. 

\subsection {Procedure summary}
\label {summary}
The principal steps can be summarized as follows.
\begin{itemize}
\item
The effect of fermionic annihilation-creation operators, $a_{\alpha}$ and $a^{\dagger}_{\beta}$,
on the system of \emph{one-electron states} is map\-ped onto $(\uparrow,\!\downarrow)$ 
states in a model system of $s\!=\!\tfrac{1}{2}$ spins (via the Jordan -- Wigner transformation).
\item
The state of each spin is represented by a qubit.
\item
Excitations in multi-electron system are then represented as qubits, 
which interact and run through a quantum circuit.
\item
The {circuit} consists of a number of basic elements (quantum gates),
arranged according to the structure of equations to solve.
\item
At the beginning, each {qubit} is prepared according 
to the starting configuration (i.e. occupation of the electron orbitals) chosen.
\item
The output of the circuit (measurement) yields the expectation value of each qubit.
It can be redirected to the input till convergence.
\item
The configuration emerging in the repetitive process, ta\-ken together 
with the (previously calculated) matrix elements, yields the physical solution
(energy and wavefunction).
\end{itemize}

\section{Results and Discussion} % =======================================
\label{case_study}
\subsection {Calculation details}

We  used the procedure outlined in previous sections to calculate the ground state 
energy (which can be straightforwardly converted into the dissociation energy) of two mole\-cules, hydrogen (H$_2$) and lithium hydride (LiH), also (that is presumably a novel result)
under the action of stationary electric fields of four different magnitudes ($\mathbb{E}=$ 0.0001, 0.001, 0.01, 0.1 atomic units; 1~a.u.~$\approx 5\cdot 10^{11}$~V/m). 
These calculations were performed for the interatomic distances, $d$, from 0.2 to 4~{\AA} with the step 0.1~{\AA}.

The actual computational environment, where these experiments were conducted, was the {\tt IBM Q},
an ensemble of quantum computers and simulators and able to perform quantum computation. Such computational environment is  available remotely through the internet and can be accessed and programmed using the {\tt QISKit} framework, written in the {\tt Python} language.
The actual code developed to this work is available in the following {\tt github} repository: 
{\small \url {https://github.com/arcalab/experiments_quantum_chemistry/tree/master/Qiskit_Programmatic_version_src}};
it makes use notably of the {\tt QISkit} and the {\tt PySCF} {\tt python} framework. 

The {\tt PySCF} tool was used to specify the molecules and calculate the respective one-body and two-body integrals,
encompassing already the action of electric fields, using the theory developed throughout Appendix \ref {appendix_A}. 
Both molecules were assumed to have zero
global charge and spin zero; the STO-3G basis \eqref{sto3g} was used to calculate the integrals.

The tasks of evaluation of corresponding integrals can then be 
reformulated into an assembling of
quantum circuits, to be executed in quantum computers supplied,
using the set of \emph {software} packages available 
e.g. in the {\tt QISkit} framework: {\tt Terra}, {\tt Aer}, {\tt Aqua} and {\tt Ignis}. 
The calculation of the dissociation curves requires the calculation of the {ground state} energies
(discussed in section \ref{vqe_method}) over a range of distances, to be able to identify the minimum (bound molecule)
and the asymptotics (separated atoms).
For this purpose we used two methods: the \textit{Exact Eigensolver} 
(classical \emph{matrix-multiplication} method,
as a benchmark) and the VQE.

We used the UCC (discussed in section \ref{sec-ground}) as the \emph{variational method}, i.e. the technique to build the \emph{ans\"atze}
for the molecules under study, and the HF approximation to obtain the initial solution for the VQE method.
In this relation, several parameters had to be considered:
the maximum number of iterations 
with the {\tt Cobyla} method,\footnote {In this quantum computation setting, an iteration in the {\tt Cobyla} method is an expensive operation in terms of computation time, and therefore one may be interested to limit the number of iterations. However, the method stops if convergence is verified and in our particular case, the method \emph {always} converged before $15000$ iterations.} the optimization level (an {\tt IBM Q} -specific parameter determining
the degree of optimization of the circuits generated), the mapping method to use, such as the Jordan-Wigner
\eqref{J-W-3}, Bravyi-Kitaev, or parity methods [see \citet{Cao-Review2019} for more information on these methods],
each offering different (precision) / (circuit size)
relationships. The technical parameters of calculation, selected after a course of trial and error, are summarized in Table~\ref{tab:calc_param}.

\begin{table}[t!]
%\label{tab:calc_param} 
\caption{\label{tab:calc_param} 
The set of technical parameters used for quantum calculations. See text for detail.}
\smallskip
\begin{tabular}{p{0.3\textwidth}p{0.12\textwidth}}
%\begin{tabular}{lc}
\hline
\rule[-4pt]{0pt}{12pt} parameter & value \\
\hline
{\tt shots}$^a$ \rule[0pt]{0pt}{9pt}  & 4096 \\
Max. number of iterations of {\tt Cobyla} & 15000 \\
Max. number of iterations of {\tt PySCF}  & 5000 \\
optimization level & 3 \\
mapping method & Jordan-Wigner \\
{\tt QISkit} version \rule[-4pt]{0pt}{5pt} & {\tt 0.13.0} \\
\hline
\end{tabular}
\\*[2pt]
$^a$number of times the execution of circuits is to be performed %\\*[-1pt]
due to the stochastic nature of quantum computers
\end{table}

The quantum or \emph {hybrid} (such as VQE) procedures in the {\tt IBM Q} require that a \emph{backend} is specified, i.e. an actual processing node able to execute the quantum circuits, 
which can be either a classical computer able to perform the quantum computation (simulator), with or without \emph{simulated quantum noise}, or a real quantum device, with a number of qubits from 2 to 53.
The results of this work were obtained using a \emph {simulator}, the \emph {qasm\_simulator}.  

\subsection {Results: H$_2$ molecule}
%-------------    OR ---------------------
\begin{figure}[b!]
\includegraphics[scale=0.30]{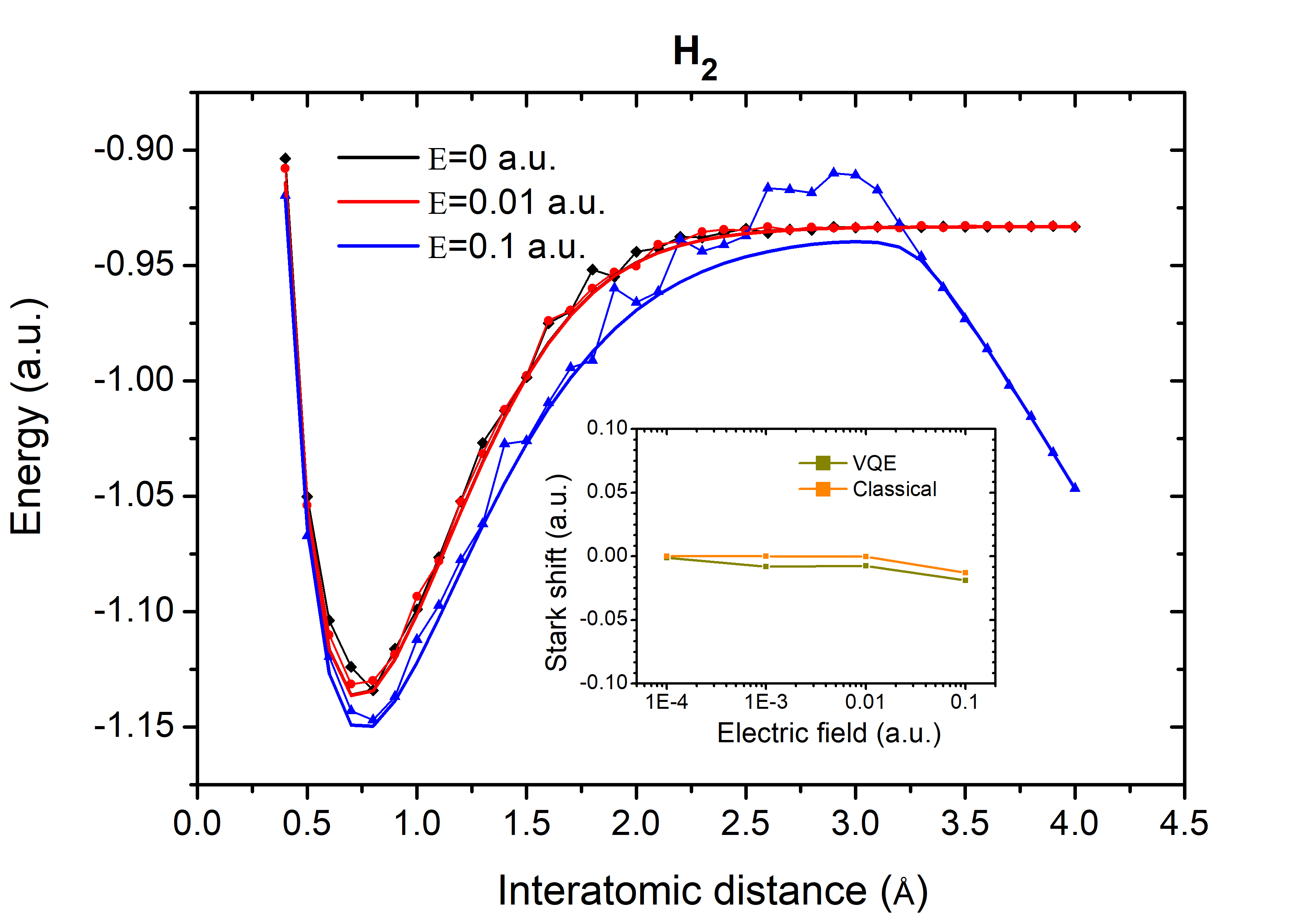}
%\smallskip
\caption{Dissociation curve of H$_2$ molecule, as calculated with a classical solver (full lines) and with the VQE (symbols connected by lines), for several values of the external electric field $\mathbb E$ marked by color.
The Stark effect (i.e. the shift of the minimum energy with electric field) is shown in the inset.
}
\label{vs:h2final_symbol}
\end{figure}

\noindent The total energy as a function of the interatomic distance,
hence the molecule's dissociation curve for different values of the electric field, 
is depicted in Figure~\ref{vs:h2final_symbol}. The effect of electric field on the shape
of the dissociation curve remains negligible at small values of the field inspected
yet results in a drastic change of the $d{\rightarrow}\infty$ asymptotic (slope)
and in a noticeable shift of the equilibrium position for $\mathbb E=0.1$~a.u. The abrupt change in the $E(d)$ dependence slope at large distances, for very large electric field $\mathbb E=0.1$~a.u., can be related with the onset of the molecule's dissociation, which becomes possible via tunneling through the energy barrier (blue curves in Fig.~\ref{vs:h2final_symbol}).

%
%\textcolor{red}{In Figure \ref{vs:h2final} are depicted the results obtained for the \emph{Hydrogen} molecule,
%where is possible to observe the effects of an electric field in the molecule's dissociation curve.
%Is very noticeable that the stronger the electric field applied, the bigger the changes obtained in the curve.} 

%\noindent 
%\textcolor{red}{As mentioned before, there were applied four different types of electrical fields ($0.0001, 0.001, 0.01$ and $0.1$) but in Figure \ref{vs:h2final} are only represented two of them.
%Small electric fields have tiny effects in the dissociation curve, and because it would turn the final data representation too clustered, it was opted only to show, beyond the $0$ electrical field, the two curves affect by the strongest fields.} 

%\noindent 
%\textcolor{red}{Furthermore, when analysing the \emph{ Classical} results (in full lines), the dissociation curve subjected to the $0.01$ electrical field (the second but still a strong electrical field) almost coincides to the $0$ electric field applied, stating that only a very strong field has a visible effect on the $H_2$ molecule's dissociation curve. When comes to the effects of the $0.1$ electrical field, is visible a shift in the energy minimum  as well as a abrupt slope in the final distance of the molecule's dissociation curve. The former can be explained as the start of the molecule's dissociation.}
%\textcolor{green}{-- Sorry, too many words. I tried to say all this differently yet shorter at the beginning of the Subsection. Please check...}

The inspection of the {VQE} results, represented by symbols connected by lines in Fig.~\ref{vs:h2final_symbol}, reveals a numerical noise that apparently increases with the electric field magnitude.
Possibly, the HF approximation used as input for the quantum calculation becomes unstable under the action of a strong electric field. 

%\begin{figure}[ht!]
% \includegraphics[scale=0.3]{images/H2_final.png}
% \caption{Dissociation curve of $H_{2}$ molecule. Results obtained with a classical solver (full line) and with the VQE (dashed %line) displayed by color according to the respective electric field applied.}
% \label{vs:h2final}
%\end{figure}

%\noindent 
The inset of Figure \ref{vs:h2final_symbol} shows the Stark effect for the mole\-cule under study, that is, the shift between the ground-state energy calculated under the action of the electric field and at $\mathbb E=0$.  The distance at which the respective energies have been extracted
was the energy minimum position yielded by the classical solver at $\mathbb E=0$, $d_{eq}=0.7$~\AA. We took this option because of the fluctuations of $E(d)$ obtained with the quantum solver.

For a non-polar molecule without intrinsic dipole moment, as is the case for H$_2$, the stationary electronic Stark effect should be quadratic in the electric field. However, with the limited minimal basis used, it looks even weaker and becomes noticeable only for very strong fields.

%\subsection{$H_{2}^{+}$ molecule}

%\begin{figure}[ht!]
% \includegraphics[scale=0.3]{images/H2+_final.png}
% \caption{Dissociation curve of $H_{2}^{+}$ molecule. Results obtained with a classical solver (full line) and with the VQE (dashed line) displayed by color according to the respective electric field applied.}
% \label{vs:h2+final}
%\end{figure}

\subsection{Results: LiH molecule}

\begin{figure}[b!]
 \includegraphics[scale=0.30]{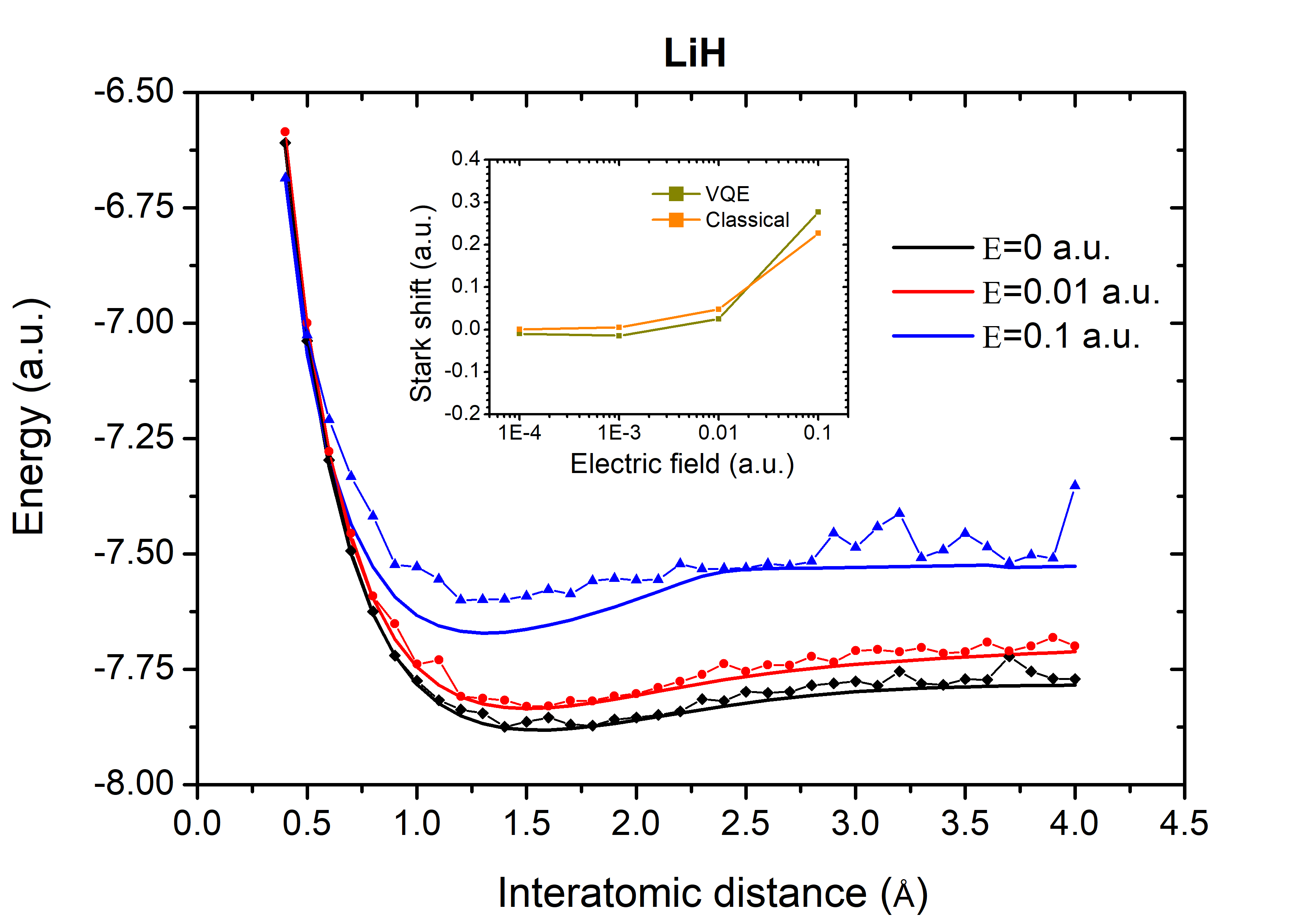}
 \caption{%
 Same as in Fig.~\ref{vs:h2final_symbol} for the LiH molecule. 
 }
\label{vs:lihfinal_symbol}
\end{figure}

The results for the lithium hydride molecule are shown in Fig.~\ref{vs:lihfinal_symbol}, where the effect of the applied electric field is quite noticeable. The displacement of the $E(d)$ curve increases with the electric field: already for 0.01 a.u. the shift of the dissociation curve becomes appreciable. The Stark effect (inset in Fig.~\ref{vs:lihfinal_symbol}) increases with the field magnitude much faster than for the H$_2$ molecule. This is because of the intrinsic dipole moment the LiH molecule already possesses in the ground state.
%because of its asymmetry, in contrast with the hydrogen molecule. 
The Stark effect is linear in $\mathbb E$ for small fields but then starts growing much faster because of the additional polarization of the ground state induced by the external field.   

Similar to the case of H$_2$ molecule, the numerical noise is visible in the results and becomes more pronounced in stronger electrical fields. Also, the ground state energy obtained with the different solvers results in different values of the equilibrium distance, 
$d_{eq}$, obtained for the quantum and classical solver, -- 1.5~{\AA} and
1.6~{\AA}, respectively, -- at $\mathbb E =0$. Again, the latter was taken as the reference value for the Stark effect evaluation.

%\begin{figure}[ht!]
% \includegraphics[scale=0.29]{images/LiH_final.png}
% \caption{Dissociation curve of $LiH$ molecule. Results obtained with a classical solver (full line) and with the VQE (dashed %line) displayed by color according to the respective electric field applied.}
% \label{vs:lihfinal}
%\end{figure}

\section {Conclusions}

We attempted to outline, in a concise way yet indicating
the essential elements and the underlying theory, a representative practical resolution 
of a simple Quantum Chemistry problem on a quantum computer. Special attention has been paid
to the connection between the fermionic Hamiltonians and the quantum circuits, as well as the state preparation, running of the algorithm and  evaluation of the results. An interested reader may wish to find out more details and discussions in the excellent recent 
review by \citet{Cao-Review2019}.
In practical terms, we programmed and executed the calculation of ground-state energies of molecules (H$_2$ and LiH),
on the commercially available (since recently) quantum computer, {\tt IBM Q}, of which we used the quantum device simulator.

The calculated results comprise the total energy as a function of bond length
(i.e. the dissociation curve), also under applied stationary
electric field. We also evaluated the shift of the molecule's energy at a fixed $d$ (equal to the equilibrium interatomic distance) with the electric field, i.e. the stationary electronic Stark effect, supposedly quad\-ra\-tic in $\mathbb E$ and small for the non-polar H$_2$ molecule but
containing the linear term and much stronger in case of the polar LiH molecule.
The quantum calculations were characterized by a considerable numerical noise, the magnitude of which
increases with the strength of the electric field. The nature of these instabilities is still
under inspection. In total,
our case study seems to provide evidence for the feasibility of the use of this quantum computer
for small molecules, with a reasonable number of iterations
performed. Thus, the current quantum computation and simulation technology, even though yet far 
from being able to address large molecules in order to answer relevant questions
in Chemistry and Biology, 
already is able to provide physically meaningful results for small systems, constituting an important milestone for further work.

\begin{acknowledgements}

The authors wish to thank Lu{\'{\i}}s Barbosa for 
helpful discussions and 
for his suggestions during the course of this work,
as well as the students of Physics Engineering at the
University of Minho, 
-- Carolina Alves, Daniel Carvalho, Michael de Oliveira and Paulo Ribeiro, -- for their helpful contributions at the preliminary stage of this work.
Carlos Tavares was funded by the FCT -- Funda{\c{c}}{\~a}o para a Ci{\^e}ncia e Tecnologia (FCT) by the grant SFRH/BD/116367/2016, funded under the POCH programme and MCTES national funds.
This work was also funded by the project ``SmartEGOV: Harnessing
EGOV for Smart Governance (Foundations, Methods, Tools) /
NORTE-01-0145-FEDER-000037'', supported by Norte Portugal
Regional Operational Programme (NORTE 2020), under the
PORTUGAL 2020 Partnership Agreement, through the European
Regional Development Fund (EFDR). Funding from the FCT in the framework of the Strategic Funding UID/FIS/04650/2019 is also gratefully acknowledged.
\end{acknowledgements}

%\section*{Compliance with ethical standards}
%
%\paragraph {\textbf {Conflict of interest}} The authors declare that they have no conflict of interest.

%\paragraph {\textbf {Ethical aproval}} This article does not contain any studies with human participants or animals performed by any of the authors.

\bibliographystyle{spbasic}

%\bibliographystyle{plainnat}% basic style, author-year citations
%\bibliography{references}   % name your BibTeX data base

\appendix
\newpage
\section{\label{appendix_A} Calculation of the matrix elements}

\subsection{STO-LG wavefunctions}

The STO-3G type combinations of Gaussian functions are used to calculate
the matrix elements of various electronic interactions in the mole\-cules under study.
As the minimal basis of the H$_2$ molecule includes the $s$-type orbitals only, 
whereas that for LiH comprises both the $s$- and the $p$-type orbitals, 
by throughout covering 
%exposing the wavefunctions for LiH we shall consider 
the latter molecule we leave a possibility to fall back to
%, which can be easily adopted for 
the H$_2$ case
%hydrogen molecule 
by removing the factor of 3 (Li nucleus charge) in those matrix elements where it appears explicitly (namely, in Table \ref {t5} below). Also, the parameters of the STO-3G functions have to be chosen accordingly (see Table \ref {tablevs:1} below).    

\begin{figure}[b!]
	\centerline{\includegraphics[width=0.6\linewidth]{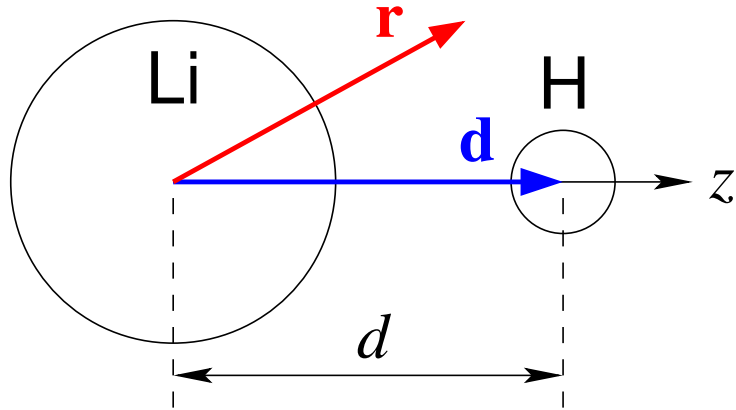}}
	\smallskip
	\caption{\label{fig:Li-H_geom}
		Geometry setting in calculations concerning the LiH molecule.}
\end{figure}

%\ref{Scheme of the LiH molecule's axis}.
%\begin{figure}[ht!]
% \centering
% \begin{tikzpicture}[x=0.5cm,y=0.5cm,z=0.3cm,>=stealth]
%% The axes
%%\draw[->] (xyz cs:z=-3) -- (xyz cs:z=5) node[above] {$y$};
%\draw[->] (xyz cs:x=0) -- (xyz cs:x=10) node[right] {$z$};
%%\draw[->] (xyz cs:y=-0) -- (xyz cs:y=5) node[above] {$x$};
%
%\draw[->] (0,0) -- (7,0) node[above] {$\vec{d}$};
%
%\draw[->] (3,0) coordinate (a) node[right] {}
%      -- (0,0) coordinate (b) node[left] {}
%      -- (3.5,1.5) coordinate (c) node[above] {$\vec{r}$}
%      pic["$\theta$", draw=black, <->, angle eccentricity=1.2, angle radius=1.5cm]
%      {angle=a--b--c};
%
%\draw[<->] (0,-3) -- (7,-3) node[below] {d};
%\draw[dashed] (0,0) -- (0,-3);
%\draw[dashed] (7,0) -- (7,-3);      
%
%\draw (0,0) circle (1cm) (0,2) node[above] {Li};
%\node at (0,0) [circle,fill,inner sep=1pt] {};
%\draw (7,0) circle (0.5cm) (7,1) node[above] {H};
%\node at (7,0) [circle,fill,inner sep=1pt] {};
%\end{tikzpicture}
%\caption{Scheme of the LiH molecule's axis
%\textcolor{green}{Replace this assembled figure by a single .eps figure}
%}
%\label{Scheme of the LiH molecule's axis}
%\end{figure}

The minimal basis will include the following atomic orbitals: $1s$ for H; $1s$, $2s$ and $2p_z$ for Li. All of them will be approximated by the STO-3G type combinations of the following Gaussian functions \citep{szabo2012modern}:
\begin{eqnarray}
\psi_{1s}(\zeta) &=& \sum\limits_{i=1}^{L} d_{i,1s} g_{1s}(\alpha_{i,1s})\,; \label{vs:1} \\
\psi_{2s}(\zeta) &=& \sum\limits_{i=1}^{L} d_{i,2s} g_{1s}(\alpha_{i,2sp})\,; \label{vs:2} \\
\psi_{2p_z}(\zeta) &=& \sum\limits_{i=1}^{L} d_{i,2p} g_{2p_z}(\alpha_{i,2sp})\,.\label{vs:3}
\end{eqnarray}
Here $\zeta$ is a parameter appearing in the Slater-type orbitals ($\zeta=1.24$
for H and $\zeta=2.69$ as the ``recommended'' value for Li$1s$); 
the coefficients $d_{i}$ and $\alpha_{i}$ are fitted parameters and $g$ are the normalized Gaussian functions:
\begin{eqnarray}
g_{s}(\alpha) &=& c_{\alpha}^{(s)}e^{-\alpha r^2},
\quad\quad\quad\quad\quad  \mbox{with} 
\quad c_{\alpha}^{(s)}=\left(\!\frac{2\alpha}{\pi}\!\right)^{\!\!\tfrac{3}{4}}\,; 
\label{vs:4} \\
g_{p}(\beta) &=& c_{\beta}^{(p)}r\cos{\theta}e^{-\beta r^2},
\quad\quad  \mbox{with} 
\quad c_{\beta}^{(p)}= \left(\!\frac{128\,\beta^5}{\pi^3}\!\right)^{\!\!\tfrac{1}{4}}\,.
\label{vs:5}
\end{eqnarray}
The fitted Gaussian exponents and the corresponding coefficients $d_{i}$ depend on the parameter $\zeta$ 
in the Slater orbital, also called ``scaling factor'', which is different for each atomic shell 
(e.g for $2s$ and $2p$ states of Li the recommended value is $\zeta=0.75$).
The exponents for $\zeta=1$ are given in Table~3.7 of \citet{szabo2012modern};
for $\zeta\neq 1$ they scale as $\alpha(\zeta)=\alpha(1)\cdot\zeta^{2}$, 
whereby the coefficients $d$ are the same for each type of states in different atoms, -- e.g $1s$ (H) and $1s$ (Li), --
although $\alpha$'s are different. The parameters used by us are compiled in Table \ref{tablevs:1}.

%\begin{table*}[t!]
%\caption{\label{tablevs:1} Values of the parameters of STO-3G orbitals for H and Li atoms}
%\centering
% \begin{tabular}{c|c|c|c|c|c}
% \hline
% \multicolumn{2}{c|}{$\alpha_{1s}$} & $d_{1s}$ & \multirow { $\alpha_{2,sp}(\text {Li},\, \zeta=0.75)$ }  &  \multirow %{$d_{2s}(\text {Li})$ } & \multirow { $d_{2p}(\text {Li})$ } \\ \cline{1-3} 
%  H($\zeta=1.24$) & Li($\zeta=2.69$) & $\text {H, Li}$ & \multicolumn{1}{c|}{} & \multicolumn{1}{c|}{} \\ \hline  %& %$\alpha_{2,sp}(\zeta=0.75)$ & $d_{2s}$(Li) & $d_{2p}$(Li)\\ \hline
%  3.425250914 & 16.11957475 & 0.1543289673 & 0.6362897469 & -0.09996722919 & 0.1559162750 \\ \hline
%  0.6239137298 & 2.936200663 & 0.5353281423 & 0.1478600533 & 0.3995128261 & 0.6076837186 \\ \hline
%  0.1688554040 & 0.7946504870 & 0.4446345422 & 0.04808867840 & 0.7001154689 & 0.3919573931 \\ \hline
% \end{tabular}
%\end{table*}

\begin{table*}[b!]
	\caption{\label{tablevs:1} Parameters of STO-3G orbitals for H and Li atoms used in the calculations.}
	\centering
	\begin{tabular}{r@{.}lr@{.}lcr@{.}lr@{.}lr@{.}lr@{.}lr@{.}l}
		\hline
		\multicolumn{4}{c}{H} & \rule[-3pt]{0pt}{11pt} & \multicolumn{10}{c}{Li} \\
		\cline{1-4} \cline{6-15}
		\multicolumn{2}{c}{\parbox[c]{1.8cm}{%
				\vspace*{-5pt} \begin{center} $\alpha_{1s}$\\($\zeta=1.24$) \end{center} \vspace*{-5pt}
		}} & 
		\multicolumn{2}{c}{$d_{1s}$} &&
		\multicolumn{2}{c}{\parbox[c]{1.8cm}{%
				\vspace*{-5pt} \begin{center} $\alpha_{1s}$\\($\zeta=2.69$) \end{center} \vspace*{-5pt}
		}} & 
		\multicolumn{2}{c}{$d_{1s}$} &
		\multicolumn{2}{c}{\parbox[c]{1.8cm}{%
				\vspace*{-5pt} \begin{center} $\alpha_{2s,p}$\\($\zeta=0.75$) \end{center} \vspace*{-5pt}
		}} & 
		\multicolumn{2}{c}{$d_{2s}$} & \multicolumn{2}{c}{$d_{2p}$} \\
		\cline{1-4} \cline{6-15}
		3&425250914  & 0&1543289673 & \rule[0pt]{0pt}{9pt} &
		16&11957475   & 0&1543289673 & 0&6362897469 & $-$0&09996722919 & 0&1559162750 \\
		0&6239137298 & 0&5353281423 & & 
		2&936200663  & 0&5353281423 & 0&1478600533 &    0&3995128261  & 0&6076837186 \\
		0&1688554040 & 0&4446345422 & \rule[-2pt]{0pt}{5pt} &
		0&7946504870 & 0&4446345422 & 0&04808867840 &   0&7001154689  & 0&3919573931 \\
		\hline
	\end{tabular}
\end{table*}

%Therefore we shall consider the matrix elements between the following three functions:
%\begin{align}
% |A\rangle = c_{\alpha}^{(s)}e^{-\alpha(\vec{r}-\vec{d})^{2}}; \quad
% |B\rangle_{1s\ or\ 2s} = c_{\beta}^{(s)}e^{-\beta r^{2}}; \quad
% |B\rangle_{2p} = c_{\beta}^{(p)}r\cos{\theta} e^{-\beta r^{2}}
%\end{align}

\subsection{One-electron matrix elements}
We shall use spherical coordinates with the origin at the Li atom,
as shown in Figure~\ref{fig:Li-H_geom}.
From now on, the Li atom will be denoted ``B'' and the H atom will be ``A'', and, according to
the previous section, we shall consider the matrix elements between the following three functions:
\begin{eqnarray}
|A\rangle = c_{\alpha}^{(s)}e^{-\alpha(\vec{r}-\vec{d})^{2}}\,; \quad
&& |B\rangle_{1s\ or\ 2s} = c_{\beta}^{(s)}e^{-\beta r^{2}}\,; \nonumber \\
&& |B\rangle_{2p} = c_{\beta}^{(p)}r\cos{\theta} e^{-\beta r^{2}}\,. \label{vs:6}
\end{eqnarray}

\subsubsection*{Nuclear Potential Energy Matrix Elements}

\begin{figure}[b!]
	\centering
	\includegraphics[width=0.36\textwidth]{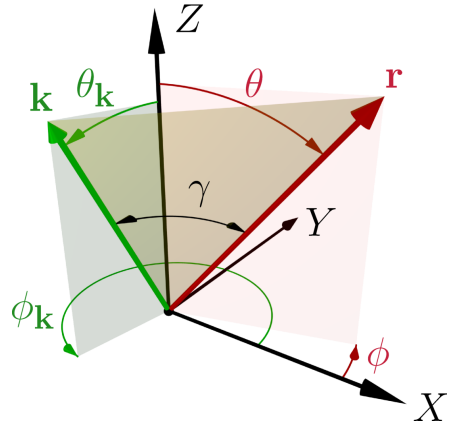}
	\smallskip
	\caption{\label{fvs:angles}
		Coordinate system and angles used in the calculation of integrals.
		See text for details.}
\end{figure}

To calculate the nuclear potential energy matrix elements, one needs to calculate the following integrals:
\begin{eqnarray}
I_{ab}^{(s(1))} &=& \langle A|\dfrac{1}{r}|B\rangle_{s} = 
c_{\alpha}^{(s)}c_{\beta}^{(s)}\int{ e^{-\alpha(\vec{r}-\vec{d})^{2}-\beta r^{2}} 
	\dfrac{1}{r}\ d\vec{r} }\,; \label{vs:7} 
\\
I_{ab}^{(s(2))} &=& \bigl\langle A\bigl|\dfrac{1}{|\vec{r}-\vec{d}|}\bigr|B\bigr\rangle_{\!\!s} =
c_{\alpha}^{(s)}c_{\beta}^{(s)}\int{ e^{-\alpha(\vec{r}-\vec{d})^{2}-\beta r^{2}} 
	\frac{1}{\vert \vec{r}-\vec{d}\vert }\ d\vec{r} }\,. 
\label{vs:8}
\end{eqnarray}
These integrals are the same as for the H$_2$ molecule, so we can use
the result of Equation (A33) from \citet{szabo2012modern}:
%
%\begin{subequations}
%
\begin{eqnarray}
I_{ab}^{(s(1))} &=& c_{\alpha}^{(s)}c_{\beta}^{(s)} \dfrac{2\pi}{\alpha+\beta} \exp{\left(-\dfrac{\alpha\beta}{\alpha+\beta}d^{2}\right)} F_{o}\left(\dfrac{\beta^{2}}{\alpha+\beta}d^{2} \right)
\label{vs:7a} \,; %\tag{59a}  
\\
I_{ab}^{(s(2))} &=& c_{\alpha}^{(s)}c_{\beta}^{(s)} \dfrac{2\pi}{\alpha+\beta} \exp{\left(-\dfrac{\alpha\beta}{\alpha+\beta}d^{2}\right)} F_{o}\left(\dfrac{\alpha^{2}}{\alpha+\beta}d^{2} \right) 
\label{vs:8a} \,,%\tag{60a}
\end{eqnarray}
%
%\end{subequations}
where 
$F_o(x)$ is expressed via the error function,
$F_{o}(x)=\sqrt{\dfrac{\pi}{4x}}\mbox{erf}\bigl(\sqrt x\bigr)$. 
The matrix elements involving the $p$-orbital are:
\begin{align}
I_{ab}^{(p(1))} &= \bigl\langle A \bigl| \frac{1}{r} \bigr| B \bigr\rangle_{\!\!p} = 
c_{\beta}^{(p)}c_{\alpha}^{(s)}\int{ e^{-\alpha(\vec{r}-\vec{d})^{2}}  
	r \cos{\theta} e^{-\beta r^{2}} \frac{1}{r}\ d\vec{r}} 
\label{vs:9} \\
&=c_{\beta}^{(p)}c_{\alpha}^{(s)}\int{ e^{-\alpha(\vec{r}-\vec{d})^{2}}  
	\cos{\theta} e^{-\beta r^{2}}\ d\vec{r} } \nonumber \\
&=c_{\beta}^{(p)}c_{\alpha}^{(s)}\int{ f_{1}(\vec{r})f_{2}(\vec{r})d\vec{r} }\,, \nonumber
\end{align}
where $f_{1}(\vec{r}) = e^{-\alpha(\vec{r}-\vec{d})^{2}}$ and $f_{2}(\vec{r})=\cos{\theta} e^{-\beta r^{2}}$.
It is convenient to use the Fourier transform of these functions:
\begin{eqnarray}
f_{1}(\vec{k}) &=& \int{ f_{1}(\vec{r})e^{-i\vec{k}\cdot\vec{r}}d\vec{r}} \nonumber \\ 
&=&  e^{-i\vec{k}\cdot\vec{d}} \int{ e^{-\alpha(\vec{r}-\vec{d})^{2}} 
	e^{-i\vec{k}(\vec{r}-\vec{d})}\ d(\vec{r}-\vec{d}) } \nonumber \\
&=&  e^{-i\vec{k}\cdot\vec{d}} \left(\!\frac{\pi}{\alpha}\!\right)^{\!\!\tfrac{3}{2}} e^{-\tfrac{k^{2}}{4\alpha}}\,;
\label{vs:10} \\ 
f_{2}(\vec{k}) &=& \int{ \cos{\theta} e^{-\beta r^{2}} e^{-i\vec{k}\cdot\vec{r}} d\vec{r} }\,. \label{vs:11}
\end{eqnarray}
For $f_2(\vec{k})$ we need to express $\cos{\theta}$ in terms of $\cos{\gamma}$, 
since %\newline
$\vec{k}\cdot\vec{r}=kr\cos{\gamma}$.
The vectors $\vec{k}$, $\vec{e_{z}}$ and $\vec{r}$, in general, do not lie in the same plane, 
so we need to consider the spherical triangle shown in the Figure~\ref{fvs:angles}. 
We can use the following formula relating the angles $\theta$, $\theta_{k}$ and $\gamma$:
\begin{align}
\cos{\theta}=\cos{\theta_{k}}\cos{\gamma}+\sin{\theta_{k}}\sin{\gamma}\cos{(\phi-\phi_{k})} \label{vs:12}
\end{align}
Using \eqref{vs:12}, we obtain:
\begin{align*}
f_{2}(\vec{k}) = 2\pi \int_{0}^{\infty} r^{2}\ dr \int_{-1}^{1} 
\cos{\theta_{k}}\cos{\gamma}e^{-\beta r^{2}-ikr\cos{\gamma}} d\gamma 
\end{align*}
[notice that the integration over $\phi$ eliminated the second term in (\ref{vs:12})]. 
The integral with respect to $\cos{\gamma}$ yields:
\begin{align*}
\int_{-1}^{1} z e^{-ikrz} dz = \dfrac{2i}{kr}\left[ \cos{(kr)} - \dfrac{\sin{(kr)}}{kr} \right] = -2ij_{i}(kr)\,,
\end{align*}
where $j_{i}(x)$ is the spherical Bessel function. Then
\begin{eqnarray}
f_{2}(\vec{k}) &=& 2\pi\cos{\theta_{k}}\left( \dfrac{2i}{k} \right) \int_{0}^{\infty} \left[ \cos{(kr)} - \dfrac{\sin{(kr)}}{kr} \right] e^{-\beta r^{2}} r\ dr 
\label{vs:13} \\
&=& 2\pi i\cos{\theta_{k}} \dfrac{1}{k\beta} \left\{ 1 - \left( \dfrac{2\beta}{k^{2}}+1 \right) F_{D}\left( \dfrac{k}{2\sqrt{\beta}} \right) \right\}\,, 
\label{vs:13a} %\tag{68a}
\end{eqnarray}
whereby
\begin{align}
F_{D}(t) = \dfrac{\sqrt{\pi}}{2} e^{-t^{2}}\mbox{erfi}(t)\,, \label{vs:14}
\end{align}
in which $\mbox{erfi}(t) = -i\mbox{erf}(t)$, is called the Dawson's function. Then
\begin{eqnarray}
I_{ab}^{(p(1))} &=& c_{\beta}^{(p)}c_{\alpha}^{(s)} \int d\vec{r} \left\{ \int f_{1}(\vec{k}_{1})
e^{i \vec{k}_{1} \vec{r}} \dfrac{d\vec{k}_{1}}{(2\pi)^{3}}  \int f_{2}(\vec{k}_{2}) e^{i \vec{k}_{2} \vec{r}} \dfrac{d\vec{k}_{2}}{(2\pi)^{3}} \right\} \nonumber \\
&=& c_{\beta}^{(p)}c_{\alpha}^{(s)} \int \dfrac{d\vec{k}}{(2\pi)^{3}} f_{2}(\vec{k})f_1(-\vec{k})\,. 
\label{vs:9a} %\tag{61a}
\end{eqnarray}

\noindent The angular part of the integral in \eqref{vs:9a} is:
\begin{align*}
2\pi \int_{-1}^{1} e^{ikd\cos{\theta_{k}}}\cos{\theta_{k}}\ d\cos{\theta_{k}} = \dfrac{4\pi}{ikd}\left[\cos{(kd)} - \dfrac{\sin{(kd)}}{kd} \right]\,, 
\end{align*}
and we have:
\begin{align}
I_{ab}^{(p(1))}(\alpha,\beta;d) &= c_{\beta}^{(p)}c_{\alpha}^{(s)} \left( \dfrac{\pi}{\alpha} \right)^{\!\!\tfrac{3}{2}} 
\nonumber\\
& \times \dfrac{1}{\pi\beta d} \int_{0}^{\infty} \Bigl[ 1 - \left( \dfrac{2\beta}{k^{2}}+1 \right) F_{D}\left( \dfrac{k}{2\sqrt{\beta}} \right) \Bigr] 
\nonumber \\
&\times \exp{\left(\!- \dfrac{k^{2}}{4\alpha}\!\right)} \left[\cos{(kd)} - \dfrac{\sin{(kd)}}{kd} \right] dk\,.
\label{vs:15}
\end{align}
Another integral of this type, describing electrons interaction with the H atom, is:
\begin{align}
I_{ab}^{(p(1))} &= \bigl\langle A \bigl| \dfrac{1}{|\vec{r}-\vec{d}|} \bigr| B \bigr\rangle_{\!\!p}
\\ 
&= c_{\beta}^{(p)}c_{\alpha}^{(s)}\int{r \cos{\theta} e^{-\beta r^{2}}
	\dfrac{1}{|\vec{r}-\vec{d}|}e^{-\alpha(\vec{r}-\vec{d})^{2}}}\ d\vec{r}\,,
\label{vs:16}
\end{align}
where $f_{1}(\vec{r})=r\cos{\theta} e^{-\beta r^{2}} $ and $f_{2}(\vec{r})=\dfrac{1}{|\vec{r}-\vec{d}|}e^{-\alpha(\vec{r}-\vec{d})^{2}}$.
\noindent The Fourier transforms of these functions are:
\begin{align}
f_1(\vec{k})&=2i\dfrac{\cos{\theta_{k}}}{k}2\pi\int_{0}^{\infty}r^{2} \left[\cos{(kr)} - \dfrac{\sin{(kr)}}{kr} \right] e^{-\beta r^{2}}\ dr 
\nonumber \\
&= i\cos{\theta_{k}}\dfrac{\pi^{\tfrac{3}{2}}k}{2\beta^{\tfrac{5}{2}}}\exp{\left(\!-\dfrac{k^{2}}{4\beta}\right)}\,;
\label{vs:17} \\
f_2(\vec{k})&= e^{-i\vec{k}\cdot\vec{d}}2\pi\int_{0}^{\infty}\int_{-1}^{1}
e^{-\alpha r^{2}-ik\cos{\gamma}} r\ dr\ d\cos{\gamma} \nonumber \\
&= \dfrac{4\pi}{k}e^{-i\vec{k}\cdot\vec{d}}\dfrac{1}{\sqrt{\alpha}}F_{D}\left(\dfrac{k}{2\sqrt{\alpha}}\right)\,.
\label{vs:18}
\end{align}
With this,
\begin{align}
I_{ab}^{(p(2))}(\alpha;\beta;d) &= c_{\alpha}^{(s)}c_{\beta}^{(p)}
\dfrac{2\pi^{\tfrac{5}{2}}}{\sqrt{\alpha}\beta^{\tfrac{5}{2}}} 
i \int \dfrac{d\vec{k}}{(2\pi)^{3}}\cos{\theta_{k}} k e^{-i\vec{k}\cdot\vec{d}}
\nonumber \\ 
&\times \exp{\left(-\dfrac{k^{2}}{4\beta}\right)}F_{D}\left(\dfrac{k}{2\sqrt{\alpha}}\right) 
\nonumber \\
&\mbox{\hspace*{-1.8cm}} = %c_{\alpha}^{(s)}c_{\beta}^{(p)} 
\frac{c_{\alpha}^{(s)}c_{\beta}^{(p)}\sqrt{\pi}}{\sqrt{\alpha}\beta^{\tfrac{5}{2}}d}\int_{0}^{\infty}\!\!\! \exp{\!\left(\!\!-\dfrac{k^{2}}{4\beta}\!\right)}F_{D}\!\left(\dfrac{k}{2\sqrt{\alpha}}\right)
\!\left[\cos{(kd)}\!-\!\dfrac{\sin{(kd)}}{kd}\!\right]kdk 
\nonumber \\*[4pt]
&\mbox{\hspace*{-1.8cm}}  = %c_{\alpha}^{(s)}c_{\beta}^{(p)}
\frac{c_{\alpha}^{(s)}c_{\beta}^{(p)}\sqrt{\pi}}{\sqrt{\alpha}\beta^{\tfrac{5}{2}}d^{3}}
\int_{0}^{\infty} (x\cos{x}-\sin{x})e^{-b'x^{2}} \, F_{D}(a'x)\ dx\,,
\label{vs:19}
\end{align}
where $b'=\dfrac{1}{4\beta d^{2}}$ and $a'=\dfrac{1}{2\sqrt{\alpha}d}$, $F_{D}$ is the Dawson's function (\ref{vs:14}).
Note that the dimension of the normalization constants is $[c_{\alpha}^{(s)}]=L^{-\tfrac{3}{2}}$,
$[c_{\beta}^{(p)}]=L^{-\tfrac{5}{2}}$, while $[\alpha]=[\beta]=L^{2}$, 
thus, overall dimension of (\ref{vs:19}) is $L^{-1}$, as it should be.
The integral in (\ref{vs:19}) couldn't be evaluated analytically, so 
it has to be calculated numerically.

We still need matrix elements of $r^{-1}$ diagonal in atomic index, which are as follows:
\begin{eqnarray}
I_{aa}^{(s(1))}(\alpha,\beta) &=& c_{\alpha}^{(s)}c_{\beta}^{(p)}\int\dfrac{1}{r}e^{-(\alpha+\beta)r^{2}}d\vec{r} 
\nonumber \\
&=& \dfrac{2\pi_{\alpha}^{(s)}c_{\beta}^{(p)}}{\alpha+\beta}; 
\label{vs:a1} \\
I_{bb}^{(s(1))}(\alpha,\beta) &=& I_{aa}^{(s(1))}(\alpha,\beta); 
\nonumber \\
I_{bb}^{(ps(1))} &=& c_{\alpha}^{(s)}c_{\beta}^{(p)}\int \dfrac{1}{r}r\cos{\theta}e^{-(\alpha+\beta)r^{2}} d\vec{r}=0\,;
\label{vs:a2} 
\end{eqnarray}

\begin{eqnarray}
I_{bb}^{(p(1))}(\alpha,\beta) &=& c_{\alpha}^{(p)}c_{\beta}^{(p)} \int \dfrac{1}{r}r^{2}\cos^{2}{\theta}e^{-(\alpha+\beta)r^{2}} d\vec{r} 
\nonumber \\
&=& \frac{4\pi}{3}\dfrac{c_{\alpha}^{(p)}c_{\beta}^{(p)}}{(\alpha+\beta)^{2}}\,;
\label{vs:a3} 
\end{eqnarray}

\begin{eqnarray}
I_{aa}^{(s(2))}(\alpha,\beta;d) &=&c_{\alpha}^{(s)}c_{\beta}^{(s)} \int \dfrac{1}{r}e^{-(\alpha+\beta)(\vec{r}-\vec{d})^{2}} d\vec{r} 
\nonumber \\
&=& c_{\alpha}^{(s)}c_{\beta}^{(s)}  \dfrac{2\pi^{\tfrac{3}{2}}}{(\alpha+\beta)^{\tfrac{3}{2}}d^2} \mbox{erf}\left( \sqrt{\alpha+\beta}\ d\right)\,;
\label{vs:a4} \\
I_{bb}^{(s(2))} &=& I_{aa}^{(s(2))}\,; 
\nonumber \\
I_{bb}^{(ps(2))}(\alpha,\beta;d) &=& c_{\alpha}^{(s)}c_{\beta}^{(p)} \int \dfrac{1}{|\vec{r}-\vec{d}|} e^{-(\alpha+\beta)r^{2}} r\cos{\theta}\ d\vec{r}\,.
\label{vs:a5}
\end{eqnarray}
Here we use the following expansion:
\begin{align}
\dfrac{1}{|\vec{r}-\vec{d}|} &=\dfrac{1}{\sqrt{r^{2}+d^{2}-2rd\cos{\theta}}} \nonumber \\ 
&= \dfrac{1}{d}\sum\limits_{l=0}^{\infty} P_{l}(\cos{\theta})\times 
\begin{cases}
x^{l} (x<1)\,, \\
x^{-l-1} (x>1)\,,
\end{cases} 
\label{vs:a6}
\end{align}
where $x=\dfrac{r}{d}$; since $\cos{\theta}=P_{1}(\cos{\theta})$ ($P_{l}$ are the Legendre polynomials),
the angular integration in \eqref{vs:a5} eliminates all the terms in the sum over $l$ except $l=1$. 
Therefore, we have:
\begin{align}
I_{bb}^{(ps(2))}(\alpha,\beta;d) &=  c_{\alpha}^{(s)}c_{\beta}^{(p)} \Bigg\{ \dfrac{1}{d^{2}} \int_{0}^{d} e^{-(\alpha+\beta)r^{2}}r^{4}\ dr \nonumber \\
& + d\int_{d}^{\infty} e^{-(\alpha+\beta)r^{2}} r\ dr \Bigg\} \times \frac{4\pi}{3} \nonumber \\
& \mbox{\hspace*{-1.8cm}}= \pi\dfrac{\sqrt{\pi}\,\mbox{erf}(\sqrt{\alpha\!+\!\beta}\ d) - 2\sqrt{\alpha\!+\!\beta}\ d\ e^{-(\alpha+\beta)d^{2}} }{2(\alpha+\beta)^{\tfrac{5}{2}}d^{2}}c_{\alpha}^{(s)}c_{\beta}^{(p)}\,.
\label{vs:a7}
\end{align}
Finally, the last integral of this type is:
\begin{align}
I_{bb}^{(p(2))}(\alpha,\beta;d)&=c_{\alpha}^{(p)}c_{\beta}^{(p)}  \int \dfrac{1}{|\vec{r}-\vec{d}|} e^{-(\alpha+\beta)r^{2}} r^{2}\cos^{2}{\theta}\ d\vec{r}\,.
\label{vs:a8}
\end{align}
Again, we use the formula \eqref{vs:a6} and the relation
\begin{align}
z^{2}=\dfrac{2}{3}\left[ P_{2}(z)+\dfrac{1}{2}P_{0}(z)\right]\,.
\label{vs:a9}
\end{align}
Using (\ref{vs:a9}), the angular integration in (\ref{vs:a8}) yields:
\begin{align*}
& 2\pi \int_{-1}^{1}\cos^{2}{\theta}\sum\limits_{l=1}^{\infty}P_{l}(\cos{\theta})\times
\left\{
\begin{array}{lc}
\left(\dfrac{r}{d}\right)^{l} & (r<d) \\*[6pt]
\left(\dfrac{r}{d}\right)^{-l-1} & (r>d) 
\end{array}  \right\}
d\cos{\theta} \\
&= \dfrac{8\pi}{15} 
\left\{
\begin{matrix}
\left(\dfrac{r}{d}\right)^{\!\!2}\\*[6pt]
\left(\dfrac{d}{r}\right)^{\!\!3}
\end{matrix}
\right\} + 
\dfrac{4\pi}{3}
\left\{
\begin{matrix}
1 \\*[6pt]
\dfrac{d}{r} 
\end{matrix} \right\}\,.
\end{align*}
The result is:
\begin{align}
I_{bb}^{(p(2))}(\alpha,\beta;d)&=\pi \Bigg\{ \dfrac{\left[1+(\alpha+\beta)d^{2} \right]\sqrt{\pi}
	\,\mbox{erf}(\sqrt{\alpha\!+\!\beta}\,d)  }{2(\alpha+\beta)^{\tfrac{7}{2}}d^{3}}  
\nonumber \\
& \quad\quad -\dfrac{\left[1+(\alpha+\beta)d^{2} \right]}{(\alpha+\beta)^{3}d^{2}} e^{-(\alpha+\beta)d^{2}} 
\Bigg\}\,. \label{vs:a10}
\end{align}

\subsection{Kinetic Energy Matrix Elements}
The calculation of the kinetic energy matrix elements involves the following integrals:
\begin{eqnarray}
K_{ab}^{(s)} &=& \langle A| -\nabla^{2}| B\rangle_{s} = -c_{\alpha}^{s}c_{\beta}^{s}\int e^{-\alpha(\vec{r}-\vec{d})^{2}} \nabla^{2} e^{-\beta r^{2}} d\vec{r} 
\label{vs:20} \\
&=& -c_{\alpha}^{s}c_{\beta}^{s}\int \dfrac{d\vec{k}}{(2\pi)^{3}} f_{1}\left(\vec{k}\right)f_{2}\left(-\vec{k}\right)\,, 
\label{vs:20a} %\tag{90a}
\end{eqnarray}
where $f_{2}(\vec{r}) = e^{-\alpha(\vec{r}-\vec{d})^{2}}$ and $f_{1}(\vec{r}) = \nabla^{2} e^{-\beta r^{2}}$.
{Fourier transforms of these functions are}:
\begin{align}
f_{1}\left(\vec{k}\right)&= \int d\vec{r}\ e^{-i\vec{k}\cdot\vec{r}} \nabla^{2} e^{-\beta r^{2}} = -k^{2} \left(\dfrac{\pi}{\alpha}\right)^{\!\!\tfrac{3}{2}}
\exp{\left(\!-\dfrac{k^{2}}{4\beta} \!\right)}\,.
\label{vs:22}\\
f_{2}\left(\vec{k}\right)&=e^{-i\vec{k}\cdot\vec{d}}\left(\dfrac{\pi}{\alpha}\right)^{\!\!\tfrac{3}{2}}
\exp{\!\left(-\dfrac{k^{2}}{4\alpha}\!\right)}; 
\label{vs:21}
\end{align}
Then 
\begin{align*}
K_{ab}^{(s)}&=c_{\alpha}^{s}c_{\beta}^{s} \left( \dfrac{\pi^{2}}{\alpha\beta} \right)^{\!\!\tfrac{3}{2}}
\int \dfrac{d\vec{k}}{(2\pi)^{3}} e^{-i\vec{k}\cdot\vec{d}} 
\exp{\left(\! -\dfrac{\alpha+\beta}{4\alpha\beta} k^{2}\!\right)} k^{2}\\
&= c_{\alpha}^{s}c_{\beta}^{s}  \dfrac{\pi}{2(\alpha\beta)^{\!\!\tfrac{3}{2}}} \int_{0}^{\infty} 
\dfrac{\sin{kd}}{kd} \exp{\left(\! -\dfrac{\alpha+\beta}{4\alpha\beta} k^{2}\! \right)} k^{4} dk\,. 
\end{align*}
Denoting $x=kd$, we have the following integral,
$\int_{0}^{\infty} \sin{x} e^{-bx^{2}}x^{3}\ dx$, where 
$b=\dfrac{\alpha+\beta}{4\alpha\beta d^{2}}$.
The result of the integration reads:
\begin{align}
K_{ab}^{(s)}=c_{\alpha}^{s}c_{\beta}^{s} \dfrac{4\pi^{\tfrac{3}{2}}(\alpha\beta)^{2}d^{2} }
{(\alpha\!+\!\beta)^{\!\tfrac{7}{2}}} \left[ \dfrac{3(\alpha\!+\!\beta)}{2\alpha\beta d^{2}}-1 \right] 
\exp{\left(\!-\dfrac{\alpha\!+\!\beta}{4\alpha\beta}d^{2}\!\right)}\,. 
\label{vs:23}
\end{align}
The similar integral involving the $s$ and $p$ states:
\begin{align}
K_{ab}^{(p)} &= -c_{\alpha}^{s}c_{\beta}^{p} \int r \cos{\theta} e^{-\beta r^{2}}\ \nabla^2 
e^{-\alpha(\vec{r}-\vec{d})^2}\ d\vec{r} 
\nonumber \\
&=-c_{\alpha}^{s}c_{\beta}^{p} \int \dfrac{d\vec{k}}{(2\pi)^{3}} f_{1}(\vec{k})f_{2}(-\vec{k})\,,
\quad\quad\mbox{with}
\label{vs:24} \\
f_{2}(\vec{k}) &= e^{-i\vec{k}\cdot\vec{d}} \int d(\vec{r}-\vec{d}) 
e^{-i\vec{k}\cdot(\vec{r}-\vec{d})}\ \nabla^{2} e^{-\alpha(\vec{r}-\vec{d})^2} 
\nonumber \\
&= e^{-i\vec{k}\cdot\vec{d}} k^{2} \left( \dfrac{\pi}{\alpha} \right)^{\!\tfrac{3}{2}} 
\exp{\!\left(- \dfrac{k^{2}}{4\alpha} \!\right)} \quad\quad \mbox{and}
\label{vs:25}\\
f_{1}(\vec{k}) &= i \dfrac{4\pi}{k} \cos{\theta_{k}} \int_{0}^{\infty} 
\left[\cos{(kr)}-\frac{\sin{(kr)}}{kr}\right] e^{-\beta r^{2}} r^{2}\ dr 
\nonumber\\
& = i \cos{\theta_{k}} \dfrac{\pi^{\!\tfrac{3}{2}}k}{2\beta^{\!\tfrac{5}{2}}}
\exp{\left(\!- \dfrac{k^{2}}{4\beta} \right)}\,. \label{vs:26}
\end{align}
Using \eqref{vs:25} and \eqref{vs:26},
\begin{align*}
K_{ab}^{(p)} &= c_{\alpha}^{s}c_{\beta}^{p} i \dfrac{\pi^{3}}{2\alpha^{\!\tfrac{3}{2}}\beta^{\!\tfrac{5}{2}}}
\int \cos{\theta_{k}} e^{i(\vec{k}\cdot\vec{d})} k^3 
\exp{\left(\!-\dfrac{\alpha+\beta}{4\alpha\beta} k^{2}\!\right)} \dfrac{d\vec{k}}{(2\pi)^{3}} \\
&= c_{\alpha}^{s}c_{\beta}^{p} \dfrac{\pi^{2}}{2\alpha^{\tfrac{3}{2}}\beta^{\tfrac{5}{2}}d^6}
\int_{0}^{\infty} \left( \cos{x} - \dfrac{\sin{x}}{x} \right) x^{4} e^{-bx^{2}} dx\,.
\end{align*}
The integral is calculated with the help of \textit{Mathematica}, with the result:
\begin{align}
K_{ab}^{(p)}&(\alpha,\beta;d) = c_{\alpha}^{s}c_{\beta}^{p} 
\frac{\pi^{2}}{64\alpha^{\tfrac{3}{2}}\beta^{\tfrac{5}{2}}d^{6}} b^5 \times \Bigg\{1+4b(3b-4) 
\nonumber \\
& \quad\quad
- [1+6b(4b^{2}+6b-3)]\dfrac{1}{\sqrt{b}}F_{D}\left(\dfrac{1}{2\sqrt{b}}\right) \Bigg\}\,,
\label{vs:27}
\end{align}
where $b=\dfrac{\alpha+\beta}{4\alpha\beta d^{2}}$ and $F_{D}$ is the Dawson's function (\ref{vs:14}).
The matrix elements diagonal in atomic index are as follows:
\begin{align}
K_{aa}^{(s)}(\alpha,\beta) &= c_{\alpha}^{(s)}c_{\beta}^{(s)}
\left( \dfrac{\pi^{2}}{\alpha\beta} \right) \dfrac{1}{2\pi^{2}} \int_{0}^{\infty} 
\exp{\left(- \dfrac{\alpha+\beta}{4\alpha\beta}k^{2} \right)\ k^{4}\ dk} \nonumber \\
&= c_{\alpha}^{(s)}c_{\beta}^{(s)} \dfrac{3\pi^{\tfrac{3}{2}}}{2} \dfrac{\alpha\beta}{(\alpha+\beta)^{\tfrac{5}{2}}}\,; \label{vs:28} \\
K_{bb}^{(s)}(\alpha,\beta)&=K_{aa}^{(s)}(\alpha,\beta)\,; \nonumber \\
K_{bb}^{(sp)}(\alpha,\beta) &= 0 \qquad \mbox{by symmetry}\,; \nonumber \\
K_{bb}^{(p)}(\alpha,\beta)&= \langle B |_{p} -\nabla^{2} |B\rangle_{p} \nonumber \\
& \mbox{\hspace*{-1.2cm}}
= -c_{\alpha}^{(p)}c_{\beta}^{(p)}\!\int\! \dfrac{d\vec{k}}{(2\pi)^{3}} \!
\exp{ \left(\!\!-\dfrac{\alpha\!+\!\beta}{4\alpha\beta}\!\right)k^{2} } 
\left(\!-k^{2}\cos^{2}{\theta_{k}}\dfrac{\pi^{3}k^{2}}{2(\alpha\beta)^{\tfrac{5}{2}}} \!\right) \nonumber \\
&= c_{\alpha}^{(p)}c_{\beta}^{(p)} \dfrac{5\pi^{\tfrac{3}{2}}\alpha\beta}{2(\alpha+\beta)^{\tfrac{7}{2}}}\,.
\label{vs:29}
\end{align}

\subsubsection*{Summary of one-electron Hamiltonian (for zero external field):}

\begin{table*}[t!]
	\centering
	\caption{\label{t5} %
		Matrix elements of one-electron interactions}
	%\begin{tabular}{l@{\hspace*{2mm}}c@{\hspace*{2mm}}c@{\hspace*{2mm}}c@{\hspace*{2mm}}c}
	\begin{tabular}{p{0.05\textwidth}@{\hspace*{4mm}}
			p{0.18\textwidth}p{0.18\textwidth}p{0.18\textwidth}p{0.18\textwidth}}
		\hline
		\rule[-5pt]{0pt}{15pt}
		& \multicolumn{1}{c}{$|A_{1s}\rangle $} & \multicolumn{1}{c}{$|B_{1s}\rangle $} & 
		\multicolumn{1}{c}{$|B_{2s}\rangle $} & \multicolumn{1}{c}{$|B_{2p}\rangle $} \\
		\hline
		\rule[-20pt]{0pt}{44pt}
		$\langle A_{1s}|$ &
		\parbox[c]{2.0cm}{ $K_{aa}^{(s)}(\alpha_{i,1s},\alpha_{j,1s})$
			$-I_{aa}^{(s(1))}(\alpha_{i,1s},\alpha_{j,1s})$
			$-3I_{aa}^{(s(2))}(\alpha_{i,1s},\alpha_{j,1s};d)$ } &
		\parbox[c]{2.0cm}{$K_{ab}^{(s)}(\alpha_{i,1s},\alpha_{j,1s}';d)$
			$-3I_{ab}^{(s(1))}(\alpha_{i,1s},\alpha_{j,1s}';d)$
			$-I_{ab}^{(s(2))}(\alpha_{i,1s},\alpha_{j,1s};d)$} &
		\parbox[c]{2.0cm}{$K_{ab}^{(s)}(\alpha_{i,1s},\alpha_{j,2sp}';d)$
			$-3I_{ab}^{(s(1))}(\alpha_{i,1s},\alpha_{j,2sp}';d)$
			$-I_{ab}^{(s(2))}(\alpha_{i,1s},\alpha_{j,2sp}';d)$} &
		\parbox[c]{2.0cm}{$K_{ab}^{(p)}(\alpha_{i,1s},\alpha_{j,1s}';d)$
			$-3I_{ab}^{(p(1))}(\alpha_{i,1s},\alpha_{j,2sp}';d)$
			$-I_{ab}^{(p(2))}(\alpha_{i,1s},\alpha_{j,2sp}';d)$} \\
		\hline
		\rule[-20pt]{0pt}{44pt}
		$\langle B_{1s}|$ &  & 
		\parbox[c]{2.0cm}{$K_{bb}^{(s)}(\alpha_{i,1s}',\alpha_{j,1s}')$
			$-3I_{bb}^{(s(1))}(\alpha_{i,1s}',\alpha_{j,1s}')$
			$-I_{bb}^{(s(2))}(\alpha_{i,1s}',\alpha_{j,1s}';d)$} &
		\parbox[c]{2.0cm}{$K_{bb}^{(s)}(\alpha_{i,1s}',\alpha_{j,2sp})$
			$-3I_{bb}^{(s(1))}(\alpha_{i,1s}',\alpha_{j,2sp})$
			$-I_{bb}^{(s(2))}(\alpha_{i,1s}',\alpha_{j,2sp};d)$} &
		\parbox[c]{2.0cm}{$-I_{bb}^{(ps(2))}(\alpha_{i,1s}',\alpha_{j,2sp};d)$} \\
		\hline
		\rule[-20pt]{0pt}{44pt}
		$\langle B_{2s}|$ &  &  & 
		\parbox[c]{2.0cm}{$K_{bb}^{(s)}(\alpha_{i,2sp},\alpha_{j,2sp})$
			$-3I_{bb}^{(s(1))}(\alpha_{i,2sp},\alpha_{j,2sp})$
			$-I_{bb}^{(s(2))}(\alpha_{i,2sp},\alpha_{j,2sp};d)$} &
		\parbox[c]{2.0cm}{$-I_{bb}^{(ps(2))}(\alpha_{i,2sp},\alpha_{j,2sp};d)$} \\
		\hline
		\rule[-20pt]{0pt}{44pt}
		$\langle B_{2p}|$ &  &  &  & 
		\parbox[c]{2.0cm}{$K_{bb}^{(p)}(\alpha_{i,2sp},\alpha_{j,2sp})$
			$-3I_{bb}^{(p(1))}(\alpha_{i,2sp},\alpha_{j,2sp})$
			$-I_{bb}^{(p(2))}(\alpha_{i,2sp},\alpha_{j,2sp};d)$}\\
		\hline
	\end{tabular}
\end{table*}

The one-electron Hamiltonian in the absence of external electric field is as follows:
$$
H_{1}=-\nabla^{2}-\dfrac{3}{r}-\dfrac{1}{|\vec{r}-\vec{d}|}\,.
$$
For convenience, the necessary integrals are presented in Table \ref{t5}, and Table \ref{t6} indicates the reference of the corresponding equation.

\begin{table}[t!]
	\caption{\label{t6}
		Equations specifying matrix elements of Table \ref{t5}.}
	\centering
	\begin{tabular}{ccccc}
		\hline
		Matrix & && Matrix & \\*[-1pt]  elements & Eq. && elements & Eq. \\
		\hline
		$K_{aa}^{(s)}(\alpha,\beta)$  & \eqref{vs:28}  & \rule[-2pt]{0pt}{13pt} & 
		$I_{aa}^{(s(1))}(\alpha,\beta)$ & \eqref{vs:a1} \\
		
		$K_{ab}^{(s)}(\alpha,\beta;d)$ & \eqref{vs:23} & \rule[-2pt]{0pt}{13pt} & 
		$I_{aa}^{(s(2))}(\alpha,\beta;d)$ & \eqref{vs:a4} \\
		
		$K_{ab}^{(p)}(\alpha,\beta;d)$ & \eqref{vs:27}  & \rule[-2pt]{0pt}{13pt} & 
		$I_{ab}^{(s(1))}(\alpha,\beta;d)$ & \eqref{vs:7a} \\
		
		$K_{bb}^{(s)}=K_{aa}^{(s)}$ & & \rule[-5pt]{0pt}{16pt} & 
		$I_{ab}^{(s(2))}(\alpha,\beta;d)$  & \eqref{vs:8a} \\
		
		$K_{bb}^{(p)}(\alpha,\beta;d)$  & \eqref{vs:29}  & \rule[-2pt]{0pt}{13pt} & 
		$I_{ab}^{(p(1))}(\alpha,\beta;d)$ & \eqref{vs:15} \\
		
		$I_{ab}^{(p(2))}(\alpha,\beta;d)$ & \eqref{vs:19} & \rule[-2pt]{0pt}{13pt} & 
		$I_{bb}^{(s(1))}=I_{aa}^{(s(1))}$ &  \\
		
		$I_{bb}^{(s(2))}=I_{aa}^{(s(2))}$ &  & \rule[-2pt]{0pt}{13pt} & 
		$I_{bb}^{(ps(2))}(\alpha,\beta;d)$ & \eqref{vs:a7} \\
		
		$I_{bb}^{(p(1))}(\alpha,\beta)$ & \eqref{vs:a3}  & \rule[-5pt]{0pt}{16pt} & 
		$I_{bb}^{(p(2))}(\alpha,\beta;d)$  & \eqref{vs:a10} \\
		\hline
	\end{tabular}
\end{table}

%\begin{table}[t!]
%\centering
%\renewcommand{\arraystretch}{1.5}
% \begin{tabular}{c|c}
%  $K_{aa}^{(s)}(\alpha,\beta)$ & $Eq. \ref{vs:28}$ \\ \hline
%  $I_{aa}^{(s(1))}(\alpha,\beta)$ & $Eq. \ref{vs:a1}$\\ \hline 
%  $K_{ab}^{(s)}(\alpha,\beta;d)$ & $Eq. \ref{vs:23}$ \\  \hline
%  $I_{aa}^{(s(2))}(\alpha,\beta;d)$ & $Eq. \ref{vs:a4}$ \\ \hline
%  $K_{ab}^{(p)}(\alpha,\beta;d)$ & $Eq. \ref{vs:27}$  \\ \hline
%  $I_{ab}^{(s(1))}(\alpha,\beta;d)$ & $Eq. \ref{vs:7a}$ \\ \hline
%  $K_{bb}^{(s}=K_{aa}^{(s)}$ & \\ \hline
%  $I_{ab}^{(s(2))}(\alpha,\beta;d)$ & $Eq. \ref{vs:8a}$ \\ 
% \end{tabular}
% \qquad\qquad\qquad\qquad
% \begin{tabular}{c|c}
%  $K_{bb}^{(p)}(\alpha,\beta;d)$ & $Eq. \ref{vs:29}$  \\ \hline
%  $I_{ab}^{(p(1))}(\alpha,\beta;d)$ & $Eq. \ref{vs:15}$ \\ \hline
%  $I_{ab}^{(p(2))}(\alpha,\beta;d)$ & $Eq. \ref{vs:19}$ \\\hline
%  $I_{bb}^{(s(1))}=I_{aa}^{(s(1))}$ & \\\hline
%  $I_{bb}^{(s(2))}=I_{aa}^{(s(2))}$ & \\\hline
%  $I_{bb}^{(ps(2))}(\alpha,\beta;d)$ & $Eq. \ref{vs:a7}$ \\ \hline
%  $I_{bb}^{(p(1))}(\alpha,\beta)$ & $Eq. \ref{vs:a3}$ \\\hline
%  $I_{bb}^{(p(2))}(\alpha,\beta;d)$ & $Eq. \ref{vs:a10}$
% \end{tabular}
% \caption{Relations between matrix elements of Table \ref{Matrix elements of the %one-electron Hamiltonian without electric field} and equations}
% \label{t2}
%\end{table}

\subsection{Matrix elements of the interaction with external electric field}
We shall consider the field parallel to the $z$ axis, so the interaction Hamiltonian reads:
\begin{align*}
H_S = -\mathbb{E}r\cos{\theta}\,.
\end{align*}
We shall keep the same notation as for the kinetic energy matrix elements just changing $K\rightarrow J$. First, we have:
\begin{align}
J_{aa}=\mathbb{E}d; \quad\quad J_{bb}^{(pp)}=J_{bb}^{(ss)}=0\,, \label{vs:30}
\end{align}
because the diagonal matrix elements for any atom vanish for non-degenerate atomic states and $J_{aa}$ is compensated by the energy of the proton at point $\vec{d}$ (see Fig.~\ref{fig:Li-H_geom}). For the matrix element between the $s$ and $p$-orbitals of the Li atom we have:
\begin{align}
J_{bb}^{(ps)}=J_{bb}^{(sp)}&=-c_{\alpha}^{(s)}c_{\beta}^{(p)}\mathbb{E}\int r^{2}\cos^{2}{\theta} e^{-(\alpha+\beta)r^{2}}\ d\vec{r} \nonumber \\ 
&= c_{\alpha}^{(s)}c_{\beta}^{(p)}\mathscr{E} \dfrac{\pi^{\tfrac{3}{2}}}{2(\alpha+\beta)^{\tfrac{5}{2}}}\,.
\label{vs:31}
\end{align}
The matrix elements $J_{ab}^{(s)}$ are the same as for H$_{2}$:
\begin{align}
J_{ab}^{(s)}(\alpha,\beta;d)=-c_{\alpha}^{(s)}c_{\beta}^{(s)}\int \left(\vec{\mathscr{E}}\cdot\vec{r} \right) e^{ -\alpha(\vec{r}-\vec{d})^{2} -\beta r^{2}} d\vec{r}\,. 
\label{vs:32}
\end{align}
We use the transformation:
\begin{align}
e^{ -\alpha(\vec{r}-\vec{d})^{2} -\beta r^{2}} = 
\exp \left( -\dfrac{\alpha\beta}{\alpha+\beta}\,d^2 \right) e^{-p(\vec{r}-\vec{R}_P)}\,,
\label{vs:33}
\end{align}
where $P=\alpha+\beta$ and 
$\vec{R}_P=\dfrac{1}{p}\left( \alpha\vec{R}_A+\beta\vec{R}_B\right)=\dfrac{\alpha}{p}\vec{d}$.
Then
\begin{align*}
&J_{ab}^{(s)}=-c_{\alpha}^{(s)}c_{\beta}^{(s)}\exp{ \left(\!-\dfrac{\alpha\beta}{\alpha+\beta} d^{2}\!\right) }
\int \bigl[ \vec{\mathscr{E}}\cdot(\vec{r}-\vec{R}_P) \\ 
& \qquad\qquad
+ \vec{\mathscr{E}}\cdot\vec{R}_P\bigr] e^{-p(\vec{r}-\vec{R}_P)^2} d\vec{r}
\\
&\quad\ = -c_{\alpha}^{(s)}c_{\beta}^{(s)}\exp{ \left( -\dfrac{\alpha\beta}{\alpha+\beta} d^{2} \right) }
\left( J_{1} + J_{2} \right)\,,\quad\mbox{where} \\
J_{1} &= \int (\vec{\mathscr{E}}\cdot\vec{r}')e^{-p(\vec{r}')^{2}}d\vec{r}'=0\,, \quad\mbox{and} \\
J_{2} &= (\vec{\mathscr{E}}\cdot\vec{R}_P)\int e^{-pr^{2}}d\vec{r} = 
\left(\dfrac{\pi}{\alpha+\beta}\right)^{\!\!\tfrac{3}{2}} \dfrac{\alpha}{\alpha+\beta} \mathscr{E}d\,.
\end{align*}
Thus, we have:
\begin{equation}
J_{ab}^{(s)}(\alpha,\beta;d) = -c_{\alpha}^{(s)}c_{\beta}^{(s)}\exp{ \left( -\dfrac{\alpha\beta}{\alpha+\beta} d^{2} \right) }   \dfrac{\pi^{\tfrac{3}{2}}\alpha}{(\alpha+\beta)^{\tfrac{5}{2}}} \mathscr{E}d\,.
\label{vs:12a} % \tag{102a}
\end{equation}
%
%Notice that, for this choice of the origin (at atom B), this matrix element %does not vanish 
%for $\alpha=\beta$ as it happens for the ``symmetric choice'' at %$\vec{r}=\vec{d}/{2}$.
Obviously,
$ J_{ab}^{(s)}=J_{ba}^{(s)}$.
Now we shall calculate
\begin{align}
J_{ab}^{(p)} &=-c_{\alpha}^{(s)}c_{\beta}^{(p)}\mathscr{E} \int r^{2}\cos^{2}{\theta}e^{-\beta r^{2}} e^{-\alpha(\vec{r}-\vec{d})^{2}} d\vec{r} \label{vs:34} \\
&= -c_{\alpha}^{(s)}c_{\beta}^{(p)}\mathscr{E} \int f_{1}(\vec{r})f_{2}(\vec{r})d\vec{r}\,, \nonumber
\end{align}
where
$$
f_{1}(\vec{r})=r^{2}\cos^{2}{\theta}e^{-\beta r^{2}}
\qquad \mbox{and} \qquad 
f_{2}(\vec{r})=e^{-\alpha(\vec{r}-\vec{d})^{2}}\,.
$$
The Fourier transform of $f_{1}(\vec{r})$ is:
\begin{align*}
&f_{1}(\vec{k})= \int f_{1}(\vec{r}) e^{-i\vec{k}\cdot\vec{r}}d\vec{r}\\
&\quad\quad= \int_{0}^{\infty}dr\int_{-1}^{1}d\cos{\gamma}\int_{0}^{2\pi} d\phi \left[ \cos{\gamma}\cos{\theta_{k}}+\sin{\gamma}\sin{\theta_{k}\sin{(\phi-\phi_{k})}} \right] ^{2} \\
& \qquad\qquad \times e^{-\beta r^{2}-ikr\cos{\gamma}}\,,
\end{align*}
where we made use of \eqref{vs:12}.
The term linear in $\sin{(\phi-\phi_{k})}$ vanishes after integration over $\phi$, while $\int_{0}^{2\pi}\sin^{2}{(\phi-\phi_{k})}d\phi=\pi$. Therefore,
\begin{align*}
\int_{-1}^{1} &d\cos{\gamma} \left[ \cos^{2}{\theta_{k}}\cos^{2}{\gamma}+\dfrac{1}{2}(1-\cos^{2}{\gamma})\sin^{2}{\theta_{k}} \right] e^{-ikr\cos{\gamma}} \\
& =\dfrac{2}{(kr)^{3}} \Big\{ (3\cos^{2}{\theta_{}k}-1)kr\cos{kr} \\
& \qquad + [(1-3cos^{2}{\theta_{k}})+(kr)^{2}cos^{2}(\theta_{k})]\sin{kr} \Big\} %\\ 
\quad \equiv g(\vec{k},r)
\end{align*}
and
\begin{align}
f_{1}(\vec{k}) &= 2\pi\int_{0}^{\infty} r^{4}e^{-\beta r^{2}}g(\vec{k},r)dr \nonumber \\
&= \left( \dfrac{k^{2}}{2\beta} \right) \dfrac{1}{k^{2}} \left[ \dfrac{2\beta}{k^{2}}-\cos^{2}{\theta_{k}} \right] \left(\dfrac{\pi}{\beta} \right)^{\!\!\tfrac{3}{2}} \exp{\left( - \dfrac{k^{2}}{4\beta} \right)}; \label{vs:35}\\
f_{2}(\vec{k}) &= \int e^{-\alpha r^{2}-i\vec{k}\cdot(\vec{r}+\vec{d})} d\vec{r} = e^{i\vec{k}\cdot\vec{d}} \left( \dfrac{\pi}{\alpha} \right)^{\!\!\tfrac{3}{2}} \exp{\left(\dfrac{-k^{2}}{4\alpha} \right)}\,,
\label{vs:36}
\end{align}
The integral \eqref{vs:34} is given by
\begin{align} 
J_{ab}^{(p)} = -c_{\alpha}^{(s)}c_{\beta}^{(p)}\mathscr{E} \int f_{1}(\vec{k})f_{2}(-\vec{k}) \dfrac{d\vec{k}}{(2\pi)^{3}}\,. \label{vs:37}
\end{align}
In (\ref{vs:37}), the following angular integrals come about:
\begin{equation}
\int_{0}^{2\pi}d\phi_{k}\int_{-1}^{1} e^{ikd\cos{\theta_{k}}}d\cos{\theta_{k}} 
= 4\pi \dfrac{\sin{kd}}{kd} = 4\pi j_{0}(kd)
\label{vs:38}
\end{equation}
and
\begin{align}
\int_{0}^{2\pi}\!\!\!d\phi_{k}\int_{-1}^{1}\!\!\cos^{2}{\theta_{k}} e^{ikd\cos{\theta_{k}}}d\cos{\theta_{k}} &= 4\pi \dfrac{2kd\cos{kd}+ [(kd)^2\!-\!2]\sin{kd}}{(kd)^{3}} 
\nonumber\\
& \mbox{\hspace*{-2.6cm}} =  -4\pi\left[j_{2}(kd)+\dfrac{j_{1}(kd)}{kd}\right] \equiv -4\pi Z(kd)\,. 
\label{vs:39}
\end{align}
In (\ref{vs:38}) and (\ref{vs:39}), $j_{l}(c)$ are the spherical Bessel functions and $Z(x)$ is just a short-hand notation.
With this, Eq.~\eqref{vs:37} reduces to:
\begin{align}
J_{ab}^{(p)}&(\alpha,\beta;d)=-c_{\alpha}^{(s)} c_{\beta}^{(p)} \mathscr{E} 
\dfrac{\pi d}{\beta^{2}} \left(\dfrac{1}{4\alpha\beta d^{4}}\right)^{\!\!\tfrac{3}{2}} \nonumber\\
&\times \int_{0}^{\infty} [ 2\beta d^{2} j_{0}(x) +x^{2}Z(x)] e^{-bx^{2}} dx\,,
\label{vs:40}
\end{align}
where $b=\dfrac{\alpha+\beta}{4\alpha\beta d^{2}}$. The calculation of the integral in (\ref{vs:40}) yields:
\begin{align*}
&\sqrt{\pi} \dfrac{1+4 \left(\dfrac{\alpha+\beta}{4\alpha\beta d^{2}}\right)}{4 \left(\dfrac{\alpha+\beta}{4\alpha\beta d^{2}}\right)^{\!\!\tfrac{3}{2}}} \exp{\left( -\dfrac{1}{4 \left(\dfrac{\alpha+\beta}{4\alpha\beta d^{2}}\right)} \right)} \\
& \qquad +
\dfrac{2\beta d^{2} -2}{2}  \pi\  \mbox{erf}\left( \dfrac{1}{2\sqrt{\dfrac{\alpha+\beta}{4\alpha\beta d^{2}}}} \right) \\
&\quad = \sqrt{\pi} \dfrac{1+4b}{4 b^{\tfrac{3}{2}}} \exp{\left( -\dfrac{1}{4 b} \right)} 
+ \dfrac{a-2}{2}  \pi\  \mbox{erf}\left( \dfrac{1}{2\sqrt{ b }} \right)\,,
\end{align*}
where $a=2\beta d^{2}$.

\subsubsection*{Summary of the perturbation operator}
The matrix elements of the perturbation operator due to external electric field, $H_{S}$, are summarized in Table~\ref{tab:matri_pertu}
and the corresponding equations are referred to in Table~\ref{tab:eq_matri_pertu}.
Notice that the proton energy ($-Ed$) has been added to compensate $J_{aa}$ and it is necessary to substitute $\alpha_{i,1s}$, $\alpha_{i,2sp}$ for $\alpha$ and $\beta$, respectively and $\alpha_{i,1s}'$ is for Li in the appropriate relations.
%\ref{Matrix elements for the perturbation operator}.

\begin{table*}[t!] 
	\caption{\label{tab:matri_pertu} Matrix elements for the perturbation operator}
	\centering
	\begin{tabular}{ccccc}
		\hline
		\rule[-3pt]{0pt}{12pt}  & $|A_{1s}\rangle $ & $|B_{1s}\rangle $ & $|B_{2s}\rangle $ & $|B_{2p_z}\rangle $ \\
		\hline
		$\langle A_{1s}|$ \rule[-3pt]{0pt}{14pt} & 0 & $J_{ab}^{(s)}(\alpha_{i,1s},\alpha_{j,1s}';d)$ & 
		$J_{ab}^{(s)}(\alpha_{i,1s},\alpha_{j,2sp};d)$ & $J_{ab}^{(p)}(\alpha_{i,1s},\alpha_{j,2sp};d)$
		\\ 
		$\langle B_{1s}|$ \rule[-3pt]{0pt}{13pt} & $J_{ab}^{(s)}(\alpha_{i,1s},\alpha_{j,1s}';d)$ & 0 & 0 &
		$J_{bb}^{(ps)}(\alpha_{i,1s}',\alpha_{j,2sp};d)$
		\\ 
		$\langle B_{2s}|$ \rule[-3pt]{0pt}{13pt} & $J_{ab}^{(s)}(\alpha_{i,1s},\alpha_{j,2sp};d)$ & 0 & 0 &
		$J_{bb}^{(ps)}(\alpha_{i,2sp},\alpha_{j,2sp};d)$ 
		\\ 
		$\langle B_{2p_z}|$ \rule[-5pt]{0pt}{15pt} &  $J_{ab}^{(p)}(\alpha_{i,1s},\alpha_{j,2sp};d)$ &
		$J_{bb}^{(ps)}(\alpha_{i,1s}',\alpha_{j,2sp};d)$ & $J_{bb}^{(ps)}(\alpha_{i,2sp},\alpha_{j,2sp};d)$ & 0
		\\
		\hline
	\end{tabular}
\end{table*}

%\begin{table*}[ht!] 
% \centering
% \renewcommand{\arraystretch}{2.0}
% \begin{tabular}{c|c|c|c|c}
% \multicolumn{1}{c}{$1s\ (H),\ |A\rangle$} & \multicolumn{1}{c}{$1s\ (Li),\ 
%|B\rangle_{1s}$} & \multicolumn{1}{c}{$2s\ (Li),\ |B\rangle_{2s}$} & \multicolumn{1}{c}{$2pz\ (Li),\ |B\rangle_{2p}$} & \\
%  0 & $J_{ab}^{(s)}(\alpha_{i,1s},\alpha_{j,1s}';d)$ & $J_{ab}^{(s)}(\alpha_{i,1s},\alpha_{j,2sp};d)$ &
%\multicolumn{1}{c}{$J_{ab}^{(p)}(\alpha_{i,1s},\alpha_{j,2sp};d)$} & $1s\ (H)$\\ \cline{1-4}
%  $J_{ab}^{(s)}(\alpha_{i,1s},\alpha_{j,1s}';d)$ & 0 & 0 & \multicolumn{1}{c}{$J_{bb}^{(ps)}
%(\alpha_{i,1s}',\alpha_{j,2sp};d)$} & $1s\ (Li)$ \\ \cline{1-4}
%  $J_{ab}^{(s)}(\alpha_{i,1s},\alpha_{j,2sp};d)$ & 0 & 0 & 
%\multicolumn{1}{c}{$J_{bb}^{(ps)}(\alpha_{i,2sp},\alpha_{j,2sp};d)$} & $2s\ (Li)$\\ \cline{1-4}
%  $J_{ab}^{(p)}(\alpha_{i,1s},\alpha_{j,2sp};d)$ & $J_{bb}^{(ps)}(\alpha_{i,1s}',
%\alpha_{j,2sp};d)$ & $J_{bb}^{(ps)}(\alpha_{i,2sp},\alpha_{j,2sp};d)$ & \multicolumn{1}{c}{0} &$2pz\ (Li)$\\
% \end{tabular}
% \caption{Matrix elements for the perturbation operator}
% \label{Matrix elements for the perturbation operator}
%\end{table*}
\begin{table}[t!]
	\caption{\label{tab:eq_matri_pertu}
		Equations specifying matrix elements of Table~\ref{tab:matri_pertu}}
	\centering
	\begin{tabular}{cccccc}
		\hline
		Matrix && Matrix && Matrix & \\*[-1pt]  elements & Eq. & elements & Eq. & elements & Eq. \\
		\hline
		$J_{ab}^{s}(\alpha,\beta;d)$ & \eqref{vs:12a} & $J_{ab}^{p}(\alpha,\beta;d)$ & \eqref{vs:40} & 
		$J_{bb}^{ps}(\alpha,\beta;d)$ & \eqref{vs:31} \rule[-5pt]{0pt}{15pt} \\
		\hline
	\end{tabular}
\end{table}

%\begin{table}[ht!]
% \centering
% \renewcommand{\arraystretch}{1.5}
% \begin{tabular}{c|c}
%  $J_{ab}^{s}(\alpha,\beta;d)$ & $Eq. \ref{vs:32a}$ \\ \hline
%  $J_{ab}^{p}(\alpha,\beta;d)$ &  $Eq. \ref{vs:40}$\\ \hline
%  $J_{bb}^{ps}(\alpha,\beta;d)$ & $Eq. \ref{vs:31}$
% \end{tabular}
% \caption{Relations between matrix elements of Table \ref{Matrix elements for the perturbation operator} and equations}
% \label{t4}
%\end{table}

%\newpage

\subsection{Two-electron matrix elements}
Matrix elements of the electron--electron interaction, $r_{12}^{-1}=|\vec{r_{1}}-\vec{r_{2}}|^{-1}$, in 
the ``chemist's notation'' are written in round brackets \citep {szabo2012modern}:
\begin{align*}
(ij|r_{12}^{-1}|kl) = \int d\vec{r}_1d\vec{r}_2\psi_{i}^{*}(\vec{r}_1)\psi_{j}(\vec{r}_1)
r_{12}^{-1}\psi_{k}^{*}(\vec{r}_2)\psi_{l}(\vec{r}_2) \,,
\end{align*}
which is different from the physicist's notation for the same thing, 
$\langle ik|r_{12}^{-1}|jl\rangle $, which uses angular brackets and different order of orbitals.
Here $\psi_{i}$ denotes a molecular spatial orbital constructed as a linear combination of atomic orbitals, i.e. in our case
\begin{align}
|\psi\rangle = c_{1}|A\rangle + c_{2}|B\rangle_{1s} + c_{3}|B\rangle_{2s} + c_{4}|B\rangle_{2p}\,.
\label{vs:41}
\end{align}
The HF energy includes the so called Coulomb and exchange integrals:
\begin{align}
{J}_{ij}&=(ii|r_{12}^{-1}|jj)=\langle ij|r_{12}^{-1}|ij\rangle \qquad \mbox{(Coulomb)}\,; \label{vs:42} \\
{K}_{ij}&=(ij|r_{12}^{-1}|ji)=\langle ij| r_{12}^{-1}|ji \rangle \qquad \mbox{(exchange)}\,. \label{vs:43}
\end{align}
Since $|i\rangle$ and $|j\rangle$ are linear combinations of $g_{1s}(\vec{r}-\vec{d})$, 
$g_{1s}(\vec{r})$ and $g_{2p}(\vec{r})$ functions with different coefficients 
in the exponent, 
several kinds of integrals occur in (\ref{vs:42}) and (\ref{vs:43}), namely:
($i$) four kinds of one-center integrals; ($ii$) four kinds of two-center integrals.
We proceed by elaborating on the first type (one-center) integrals, ($i$).
\begin{align}
%(i)
D_{aa}^{(ss)}(\alpha,&\beta,\gamma,\delta) = 
\int d\vec{r}_1d\vec{r}_2 g_{1s}(\alpha,\vec{r}_1)g_{1s}(\beta,\vec{r}_1)r_{12}^{-1}
g_{1s}(\gamma,\vec{r}_2)g_{1s}(\delta,\vec{r}_2) 
\nonumber \\
&\mbox{\hspace*{-0.8cm}} = c_{\alpha}^{(s)}c_{\beta}^{(s)}c_{\gamma}^{(s)}c_{\delta}^{(s)}
\int d\vec{r}_1d\vec{r}_2\, e^{-(\alpha+\beta)r_{1}^{2}} \left(r_{12}^{-1}\right)
e^{-(\gamma+\delta)r_{2}^{2}} 
\nonumber\\
& \mbox{\hspace*{-0.8cm}}= c_{\alpha}^{(s)}c_{\beta}^{(s)}c_{\gamma}^{(s)}c_{\delta}^{(s)} \int \dfrac{d\vec{k}}{(2\pi)^{3}}
\Bigl[ \Bigl( \dfrac{\pi}{\alpha+\beta} \Bigr)^{\!\!\tfrac{3}{2}}  
\exp{\Bigl(-\dfrac{k^{2}}{4(\alpha+\beta)}\Bigr)} \Bigr] 
\nonumber \\
&\times \Bigl[ \dfrac{4\pi}{k^{2}} \Bigr] \Bigl[ \Bigl( \dfrac{\pi}{\gamma+\delta} \Bigr)^{\!\!\tfrac{3}{2}} 
\exp{\Bigl(-\dfrac{k^{2}}{4(\gamma+\delta)}\Bigr)} \Bigr] 
\nonumber\\
& \mbox{\hspace*{-0.8cm}} = c_{\alpha}^{(s)}c_{\beta}^{(s)}c_{\gamma}^{(s)}c_{\delta}^{(s)}\,
\dfrac{2\pi^{2}}{[(\alpha+\beta)(\gamma+\delta)]^{\!\!\tfrac{3}{2}}}
\dfrac{\sqrt{\pi}}{2\Bigl[\dfrac{1}{4(\alpha+\beta)}+\dfrac{1}{4(\gamma+\delta)} \Bigr]^{\!\tfrac{1}{2}}  }\,;
\nonumber \\
D_{aa}^{(ss)}(\alpha,&\beta,\gamma,\delta) = 
2\pi^{\tfrac{5}{2}} %c_{\alpha}^{(s)}c_{\beta}^{(s)}c_{\gamma}^{(s)}c_{\delta}^{(s)}
\dfrac{c_{\alpha}^{(s)}c_{\beta}^{(s)}c_{\gamma}^{(s)}c_{\delta}^{(s)} } 
%{\times}2\pi^{\tfrac{5}{2}}
{(\alpha\!+\!\beta)(\gamma\!+\!\delta)(\alpha\!+\!\beta\!+\!\gamma\!+\!\delta)^{\!\!\tfrac{1}{2}}}\,.
\label{vs:44}
\end{align}
The same expression applies to $D_{bb}^{(ss)}(\alpha,\beta,\gamma,\delta)$.
\begin{align*}
% (ii)
D_{bb}^{(pp)}&(\alpha,\beta,\gamma,\delta) = \!\!\int\!\! d\vec{r}_1d\vec{r}_2
g_{2p}(\alpha,\vec{r}_1)g_{2p}(\beta,\vec{r}_1)r_{12}^{-1}g_{2p}(\gamma,\vec{r}_2)g_{2p}(\delta,\vec{r}_2)
\\
& \mbox{\hspace*{-0.6cm}}
= c_{\alpha}^{(p)}\!c_{\beta}^{(p)}\!c_{\gamma}^{(p)}\!c_{\delta}^{(p)}\!\!\int\! d\vec{r}_1d\vec{r}_2
\cos^{2}{\theta_{1}}r_{1}^{2} e^{-(\alpha+\beta)r_{1}^{2}} \left(r_{12}^{-1}\right) \cos^{2}{\theta_{2}}r_{2}^{2}
e^{-(\gamma+\delta)r_{2}^{2}} \\
& \mbox{\hspace*{-0.6cm}}
=c_{\alpha}^{(p)}\!c_{\beta}^{(p)}\!c_{\gamma}^{(p)}\!c_{\delta}^{(p)}\!\!\int\!\! \dfrac{d\vec{k}}{(2\pi)^{3}}
\Bigg\{ 
\left[ \dfrac{k}{2(\alpha\!+\!\beta)} \right]^{2} \left[ \dfrac{2(\alpha\!+\!\beta)}{k^{2}}\!-\!\cos^{2}{\theta_{k}} \right] \\
& \mbox{\hspace*{-0.4cm}}
\times \left( \dfrac{\pi}{\alpha+\beta} \right)^{\!\!\tfrac{3}{2}} \exp{\left(\!-\dfrac{k^{2}}{4(\alpha+\beta)}\!\right)}
\Bigg\} \times \left[ \dfrac{4\pi}{k^{2}} \right] \\
& \mbox{\hspace*{-0.4cm}}
\times \Bigg\{ 
\left[ \dfrac{k}{2(\gamma+\delta)} \right]^{2}\! 
\left[ \dfrac{2(\gamma\!+\!\delta)}{k^{2}}\!-\!\cos^{2}\!{\theta_{k}} \right] 
\!\left(\!\dfrac{\pi}{\gamma\!+\!\delta}\!\right)^{\!\!\tfrac{3}{2}}
\!\!\exp{\left(\!\!-\dfrac{k^{2}}{4(\gamma\!+\!\delta)}\!\! \right)}
\Bigg\}\,,
\end{align*}
where we used the Fourier transform result \eqref{vs:35}. The calculation of such integrals finally yields:
\begin{align}
D_{bb}^{(pp)}(\alpha,&\beta,\gamma,\delta) = c_{\alpha}^{(p)}c_{\beta}^{(p)}c_{\gamma}^{(p)}c_{\delta}^{(p)}
\dfrac{\pi^{\tfrac{5}{2}}}{[(\alpha+\beta)(\gamma+\delta)]^{2}} 
\dfrac{1}{(\alpha\!+\!\beta\!+\!\gamma\!+\!\delta)^{\!\!\tfrac{1}{2}}}\nonumber \\
& \quad \times \left\{ -\dfrac{1}{12}+\dfrac{6}{5}\dfrac{(\alpha+\beta)(\gamma+\delta)}
{(\alpha+\beta+\gamma+\delta)^{2}} \right\}\,. \label{vs:45}
\end{align}

\begin{table*}[t!]
	\centering
	\caption{\label{t8} %
		Matrix elements of two-electron interactions}
	\begin{tabular}{@{\hspace*{-1mm}}p{0.05\textwidth}@{\hspace*{-0mm}}
			p{0.19\textwidth}p{0.23\textwidth}p{0.23\textwidth}p{0.23\textwidth}}
		\hline
		\rule[-5pt]{0pt}{15pt}
		& \multicolumn{1}{c}{$|A_{1s}\rangle $} & \multicolumn{1}{c}{$|B_{1s}\rangle $} & 
		\multicolumn{1}{c}{$|B_{2s}\rangle $} & \multicolumn{1}{c}{$|B_{2p}\rangle $} \\
		\hline
		\rule[-13pt]{0pt}{30pt}
		$\langle A_{1s}|$ &  $D_{aa}^{(ss)}\!(\alpha_{i,1s},\alpha_{j,1s},\alpha_{k,1s},\alpha_{l,1s})$ &
		\parbox[c]{2.0cm}{$D_{ab}^{(ssE)}\!(\alpha_{i,1s},\alpha'_{j,1s},\alpha_{k,1s},\alpha'_{l,1s};d)$,
			$D_{ab}^{(ssC)}\!(\alpha_{i,1s},\alpha_{j,1s},\alpha'_{k,1s},\alpha'_{l,1s};d)$} &
		\parbox[c]{2.0cm}{$D_{ab}^{(ssE)}\!(\alpha_{i,1s},\alpha_{j,2sp},\alpha_{k,1s},\alpha_{l,2sp};d)$,
			$D_{ab}^{(ssC)}\!(\alpha_{i,1s},\alpha_{j,1s},\alpha_{k,2sp},\alpha_{l,1sp};d)$} &
		\parbox[c]{2.0cm}{$D_{ab}^{(spE)}\!(\alpha_{i,1s},\alpha_{j,2sp},\alpha_{k,1s},\alpha_{l,2sp};d)$,
			$D_{ab}^{(spC)}\!(\alpha_{i,1s},\alpha_{j,1s},\alpha'_{k,2sp},\alpha_{l,2sp};d)$} \\
		\hline
		\rule[-13pt]{0pt}{30pt}
		$\langle B_{1s}|$ &  & $D_{bb}^{(ss)}\!(\alpha'_{i,1s},\alpha'_{j,1s},\alpha'_{k,1s},\alpha'_{l,1s})$ &
		\parbox[c]{2.0cm}{$D_{bb}^{(ss)}\!(\alpha'_{i,1s},\alpha_{j,2sp},\alpha'_{k,1s},\alpha_{l,2sp};d)$,
			$D_{bb}^{(ss)}\!(\alpha'_{i,1s},\alpha'_{k,1s},\alpha_{j,2sp},\alpha'_{l,2sp};d)$} &
		\parbox[c]{2.0cm}{$D_{bb}^{(spE)}\!(\alpha'_{i,1s},\alpha_{j,2sp},\alpha'_{k,1s},\alpha_{l,2sp};d)$,
			$D_{bb}^{(spC)}\!(\alpha'_{i,1s},\alpha'_{j,1s},\alpha_{k,2sp},\alpha_{l,2sp};d)$} \\
		\hline
		\rule[-13pt]{0pt}{30pt}
		$\langle B_{2s}|$ &  &  & $D_{bb}^{(ss)}\!(\alpha_{i,2sp},\alpha_{j,2sp},\alpha_{k,2sp},\alpha_{l,2sp})$ &
		\parbox[c]{2.0cm}{$D_{bb}^{(spE)}\!(\alpha_{i,2sp},\alpha_{j,2sp},\alpha_{k,2sp},\alpha_{l,2sp};d)$,
			$D_{bb}^{(spC)}\!(\alpha_{i,2sp},\alpha_{j,2sp},\alpha_{k,2sp},\alpha_{l,2sp};d)$} \\
		\hline
		\rule[-5pt]{0pt}{15pt}
		$\langle B_{2p}|$ &  &  &  & $D_{bb}^{(pp)}\!(\alpha_{i,2sp},\alpha_{j,2sp},\alpha_{k,2sp},\alpha_{l,2sp})$ \\
		\hline
	\end{tabular}
\end{table*}

\noindent
In the calculation of \textbf{exchange-type} integrals,
\begin{align*}
%(i&ii)
D_{bb}^{(spE)}&(\alpha, \beta,\gamma,\delta) = \\
& \mbox{\hspace*{-5mm}} = \int d\vec{r}_1d\vec{r}_2 g_{1s}(\alpha,\vec{r}_1)g_{2p}(\beta,\vec{r}_1) 
\left(r_{12}^{-1}\right) g_{1s} (\gamma,\vec{r}_2) g_{2p}  (\delta ,\vec{r}_2)  \\
& \mbox{\hspace*{-5mm}}
= c_{\alpha}^{(s)}c_{\beta}^{(p)}c_{\gamma}^{(s)}c_{\delta}^{(p)}  \int d\vec{r}_1d\vec{r}_2 
\cos{\theta_{1}}r_{1} e^{-(\alpha+\beta)r_{1}^{2}}  \dfrac{1}{r_{12}} \cos{\theta_{2}}r_{2} e^{-(\gamma+\delta)r_{2}^{2}} \\
& \mbox{\hspace*{-5mm}}
= c_{\alpha}^{(s)}c_{\beta}^{(p)}c_{\gamma}^{(s)}c_{\delta}^{(p)}\!\! \int\! \dfrac{d\vec{k}}{(2\pi)^{3}} 
\Biggl[
i\cos{\theta_{k}}\dfrac{k\pi^{\tfrac{3}{2}}}{2(\gamma+\delta)^{\!\!\tfrac{5}{2}}}
\exp{ \left(- \dfrac{k^{2}}{4(\gamma+\delta)} \right) } 
\Biggr] \\
&  \times  \left[ \dfrac{4\pi}{k^{2}} \right]\!
\Biggl[
i\cos{\theta_{k}}\dfrac{k\pi^{\!\tfrac{3}{2}}}{2(\alpha+\beta)^{\!\!\tfrac{5}{2}}}
\exp{ \left(- \dfrac{k^{2}}{4(\alpha+\beta)} \right) } 
\Biggr]\,,
\end{align*}
where we used the Fourier transform \eqref{vs:17}. The calculation of the integral finally yields:
\begin{align}
D_{bb}^{(spE)}(\alpha,\beta,\gamma,\delta) &= \pi^{\tfrac{5}{2}}\,
\dfrac{ c_{\alpha}^{(s)}c_{\beta}^{(p)}c_{\gamma}^{(s)}c_{\delta}^{(p)}}
{3(\alpha+\beta)(\gamma+\delta)(\alpha\!+\!\beta\!+\!\gamma\!+\!\delta)^{\!\!\tfrac{3}{2}}}\,. 
\label{vs:46} \quad 
\end{align}
For the \textbf{Coulomb-type} integrals,
\begin{align}
% (iv)\ \ 
D_{bb}^{(spC)}(\alpha,\beta,\gamma,\delta) &= \int d\vec{r_{1}}d\vec{r_{2}} g_{1s}(\alpha,\vec{r_{1}})g_{1s}(\beta,\vec{r_{1}})r_{12}^{-1}\nonumber \\ 
\times g_{2p}(\gamma,&\vec{r_{2}})g_{2p} (\delta,\vec{r_{2}}) \nonumber \\
=c_{\alpha}^{(s)}c_{\beta}^{(s)}c_{\gamma}^{(p)} c_{\delta}^{(p)} &\dfrac{\pi^{\tfrac{5}{2}}}{(\alpha+\beta)(\gamma+\delta)^{2}(\alpha+\beta+\gamma+\delta)^{\tfrac{1}{2}}} \nonumber \\ 
\times \Bigg[ 1-\dfrac{2}{3} & \dfrac{\alpha+\beta}{\alpha+\beta+\gamma+\delta} \Bigg]\; . 
\label{vs:47}
\end{align}
Passing now to the discussion of \textbf{two-center integrals}, we begin 
with the \textbf{exchange-type} ones, involving the $s$ functions on both centers:
\begin{align}
% (v)\ 
D_{ab}^{(ssE)}(\alpha,\beta,\gamma,\delta;d)&=\int d\vec{r}_1\,d\vec{r}_2 g_{1s}(\alpha,\vec{r}_1-\vec{d})g_{1s}(\beta,\vec{r}_1) \nonumber \\
\times \dfrac{1}{|\vec{r}_1-\vec{r}_2|} & g_{1s}(\gamma,\vec{r}_2-\vec{d})g_{1s}(\delta,\vec{r}_2) 
\label{vs:48} \\
&=\int d\vec{r}_1\,d\vec{r}_2  f_{1}(\vec{r}_1) f_{2}(r_{12}) f_{3}(\vec{r}_2)\,,
\nonumber
\end{align}
where
\begin{align*}
f_{1}(\vec{r}_1) &= g_{1s}(\alpha,\vec{r}_1-\vec{d})g_{1s}(\beta,\vec{r}_1)\, ,  \\
f_{2}(r_{12}) &= \dfrac{1}{|\vec{r}_1-\vec{r}_2|}\, , \\
f_{3}(\vec{r}_2) &= g_{1s}(\gamma,\vec{r}_2-\vec{d})g_{1s}(\delta,\vec{r}_2)\,. 
\end{align*}

Following \citet[][Appendix A]{szabo2012modern}, we first express products of Gaussian functions
occurring in $f_{1}(\vec{r}_1)$ and $f_{3}(\vec{r}_2)$ as other Gaussian. 
Normalization constants will be ignored at this step, they will be introduced in the final results. 
The integral in \eqref{vs:48} becomes:
\begin{equation}
M \int d\vec{r}_1\,d\vec{r}_2 e^{-p(\vec{r}-\vec{R}_p)^2} \dfrac{1}{r_{12}} e^{-q(\vec{r}-\vec{R}_q)^2}\,, 
\label{vs:48a} % \tag{58a}
\end{equation}
where:
\begin{align}
\vec{R}_p = \dfrac{\alpha}{p}\vec{d},\ 
\vec{R}_q=\dfrac{\gamma}{q}\vec{d},\ 
p = \alpha+\beta,\
q=\gamma+\delta\ \text{, and} \nonumber\\ 
M= \exp{\left( -\left[ \dfrac{\alpha\beta}{\alpha+\beta}+\dfrac{\gamma\delta}{\gamma+\delta} \right]d^{2} \right)}\,.
\label{vs:49}
\end{align}
Now we can use Fourier transform for each factor in the integral \eqref{vs:48a}:
\begin{align}
f_{1}(\vec{k}) &= \left(\dfrac{\pi}{p} \right)^{\!\!\tfrac{3}{2}} 
\exp{\left( -\dfrac{k^{2}}{4p}-i(\vec{k}\cdot\vec{R}_p) \right)};
\quad\quad f_{2}(\vec{k})= \dfrac{4\pi}{k^{2}}; 
\nonumber \\
f_{3}(\vec{k}) &= \left(\dfrac{\pi}{q} \right)^{\!\!\tfrac{3}{2}} 
\exp{\left(\! -\dfrac{k^{2}}{4q}-i(\vec{k}\cdot\vec{R}_q) \!\right)}\,. 
\label{vs:50}
\end{align}
The integrals over $\vec{r}_1$ and $\vec{r}_2$ introduce two $\delta$-functions of $\vec{k}$ and remove
two integrations over different $\vec{k}$-vectors that appear after substituting
the Fourier integrals into \eqref{vs:48a}, so we obtain:
\begin{align}
D_{ab}^{(ssE)}(\alpha,&\beta,\gamma,\delta;d) = 
c_{\alpha}^{(s)}c_{\beta}^{(s)}c_{\gamma}^{(s)}c_{\delta}^{(s)}\dfrac{\pi M}{2(pq)^{\tfrac{3}{2}}} 
\nonumber \\
& \qquad {\times}\int \dfrac{d\vec{k}}{k^2} \exp{ \left(\!-\dfrac{p+q}{4pq}k^2+i\vec{k}\cdot(\vec{R}_p-\vec{R}_q) \right) }  
\nonumber \\
&= c_{\alpha}^{(s)}c_{\beta}^{(s)}c_{\gamma}^{(s)}c_{\delta}^{(s)} \dfrac{\pi M}{2(pq)^{\tfrac{3}{2}}} 4\pi \int\limits_{0}^{\infty}j_{0}(kR_{z})\exp{ \left(\!-\dfrac{p+q}{4pq}k^2 \! \right) } dk  
\nonumber \\
&=c_{\alpha}^{(s)}c_{\beta}^{(s)}c_{\gamma}^{(s)}c_{\delta}^{(s)} \dfrac{\pi^{3}M}{(pq)^{\tfrac{3}{2}}} 
\mbox{erf}\left[  \sqrt{\dfrac{pq}{p+q}}\cdot R_{z}\right]\,,
\label{vs:51}
\end{align}
where $R_{z}=|R_{p}-R_{q}|$.

The \textbf{two-center} $s$-$s$ \textbf{Coulomb-type} integrals read:
\begin{align}
% (vi)\ 
D_{ab}^{(ssC)}(\alpha,\beta,\gamma,\delta;&d) = \int d\vec{r}_1 d\vec{r}_2 g_{1s}(\alpha,\vec{r}_1-\vec{d})g_{1s}(\beta,\vec{r}_1-\vec{d}) \nonumber \\ 
&\times \dfrac{1}{|\vec{r}_1-\vec{r}_2|} g_{1s}(\gamma,\vec{r}_2)g_{1s}(\delta,\vec{r}_2)\,.
\label{vs:52} 
\end{align}
We can use here the previous result with $\vec{R}_p=\vec{d}$, $\vec{R}_q=0$ \newline and 
$M\rightarrow \exp{\left( -\dfrac{\alpha\beta}{\alpha+\beta}d^{2} \right)}$.
Explicitly, we have:
\begin{eqnarray}
&& D_{ab}^{(ssC)}(\alpha,\beta,\gamma,\delta;d) = 
c_{\alpha}^{(s)}c_{\beta}^{(s)}c_{\gamma}^{(s)}c_{\delta}^{(s)} 
\dfrac{\pi^{3}}{[(\alpha+\beta)(\gamma+\delta)]^{\tfrac{3}{2}}} \nonumber
\\
&& \qquad \times
\exp{\left(\! -\dfrac{\alpha\beta}{\alpha\!+\!\beta}d^{2}\!\right)} 
\times \,\mbox{erf}\left[ \sqrt{\dfrac{(\alpha+\beta)(\gamma+\delta)}{\alpha\!+\!\beta\!+\!\gamma\!+\!\delta}}\ d \right]\,.
\label{vs:52a} % \tag{122a}
\end{eqnarray}
The \textbf{two-center} \textbf{exchange-type} integrals 
involving $s$ and $p$-functions are:
\begin{align}
%(vii)\
D_{ab}^{(spE)}(\alpha,\beta,\gamma,\delta;d) &= \int d\vec{r}_1 d\vec{r}_2
g_{1s}(\alpha,\vec{r}_1-\vec{d})g_{2p}(\beta,\vec{r}_1)
\nonumber \\ 
& \qquad \times\dfrac{1}{|\vec{r}_1-\vec{r}_2|} g_{1s}(\gamma,\vec{r}_2-\vec{d})g_{2p}(\delta,\vec{r}_2) 
\nonumber \\
&= \int d\vec{r}_1 d\vec{r}_2 f_{1}(\vec{r}_1) f_{2}(r_{12}) f_{3}(\vec{r}_2)\,,
\label{vs:53}
\end{align}
where:
\begin{align*}
f_{1}(\vec{r}_1) &= g_{1s}(\alpha,\vec{r}_1-\vec{d})g_{2p}(\beta,\vec{r}_1),\ \\
f_{2}(r_{12}) &= \dfrac{1}{|\vec{r}_1-\vec{r}_2|},\ \\
f_{3}(\vec{r}_2) &= g_{1s}(\gamma,\vec{r}_2-\vec{d})g_{2p}(\delta,\vec{r}_2)\,. 
\end{align*}
Now we shall use Fourier transform in the integral \eqref{vs:53}:
\begin{eqnarray}
f_{1}(\vec{k}) &=& \int r_{1} \cos{\theta_{1}} e^{-p(\vec{r}_1-\vec{R}_{p})^{2} - i\vec{k}\cdot\vec{r}_1} d\vec{r}_1 
\nonumber \\
&=& \dfrac{1}{R_{p}} e^{-i\vec{k}\cdot\vec{R}_{p}} 
\Biggl[
\int \bigl[(\vec{r_{1}}-\vec{R}_{p})\cdot\vec{R}_{p}\bigr] e^{ -p(\vec{r}_1-\vec{R}_{p})^2-i\vec{k}
	\cdot(\vec{r}_1-\vec{R}_{p}) } d\vec{r}_1 
\nonumber \\
&& \qquad + R_{p}^{2}\int e^{-p(\vec{r}-\vec{R}_p)^{2}-i\vec{k}\cdot(\vec{r}_1-\vec{R}_{p}) } d\vec{r}_1   
\Biggr] 
\nonumber \\
&=&  e^{-i\vec{k}\cdot\vec{R_{p}}} \Bigl[\;\;
\int r\cos{\theta}e^{-pr^{2}-i(\vec{k}\cdot\vec{r})}d\vec{r} + R_{p}\int e^{-pr^{2}}e^{-i\vec{k}\cdot\vec{r}} d\vec{r}
\Bigr] 
\nonumber \\
&=&  e^{-i\vec{k}\cdot\vec{R_{p}}}  \Biggl[
i\cos{\theta_{k}}\ \dfrac{\pi^{\tfrac{3}{2}}\ k}{2p^{\tfrac{5}{2}}} \exp{ \left(- \dfrac{k^{2}}{4p} \right) } + R_{p} \dfrac{\pi^{\tfrac{3}{2}}}{p^{\tfrac{3}{2}}} \exp{ \left(- \dfrac{k^{2}}{4p} \right) } 
\Biggr]
\nonumber \\
&& \quad\mbox{where we have used the result \eqref{vs:26}}
\nonumber \\
&=& e^{-i\vec{k}\cdot\vec{R_{p}}}\  \dfrac{\pi^{\tfrac{3}{2}}}{p^{\tfrac{3}{2}}} \Bigl[
i\cos{\theta_{k}}\ \left( \dfrac{k}{2p}\right) + R_{p} \Bigr] 
\exp{ \left(- \dfrac{k^{2}}{4p} \right) }\,.
\label{vs:54}
\end{eqnarray}
As before,
\begin{align}
f_{2}(\vec{k}) &= \dfrac{4\pi}{k^{2}}\,, \quad \mbox{and} \nonumber \\
f_{3}(\vec{k}) &= e^{-i\vec{k}\cdot\vec{R}_{q}}\  \dfrac{\pi^{\tfrac{3}{2}}}{q^{\tfrac{3}{2}}} \Bigl[
-i\cos{\theta_{k}}\! \left( \dfrac{k}{2q}\right) + R_{q}
\Bigr] 
\exp{ \left(\!- \dfrac{k^{2}}{4q}\! \right) }\,. 
\label{vs:54a} % \tag{124a}\\
\end{align}
Substituting this into \eqref{vs:53},
\begin{align}
D_{ab}&^{(spE)}(\alpha,\beta,\gamma,\delta;d) = c_{\alpha}^{(s)}c_{\beta}^{(p)}c_{\gamma}^{(s)}c_{\delta}^{(p)} 4\pi\ M\ \dfrac{\pi^{3}}{(pq)^{\!\!\tfrac{3}{2}}} 
\nonumber \\
& \quad \times  \int \dfrac{d\vec{k}}{(2\pi)^{3}} \Bigl\{ 
\Bigl[ R_{p} + i\cos{\theta_{k}}\ \Bigl(\dfrac{k}{2p}\Bigr) \Bigr] \nonumber \\
& \quad \times \Bigl[  R_{q}-  i\cos{\theta_{k}} \  \Bigl(\dfrac{k}{2q}\Bigr) \Bigr] \dfrac{1}{k^{2}}
\exp{ \Bigl[\Bigl(\!-\dfrac{1}{4q}\!+\!\dfrac{1}{4p}\!\Bigr)k^{2}\Bigr] }
\Bigr\} 
\label{vs:55} \\
&= c_{\alpha}^{(s)}c_{\beta}^{(p)}c_{\gamma}^{(s)}c_{\delta}^{(p)} M \dfrac{\pi}{2(pq)^{\tfrac{3}{2}}} 
\nonumber \\
& \quad \times \int \dfrac{d\vec{k}}{k^{2}} \Bigl[ R_{p}R_{q}+i \Bigl( \dfrac{R_{q}}{p}-\dfrac{R_{p}}{q} \Bigr) \dfrac{k}{2}\cos{\theta_{k}} + \dfrac{k^{2}}{4pq}\cos^{2}{\theta_{k}} \Bigr]  
\nonumber \\
& \quad \times  \exp{ \Bigl(\!-\dfrac{p+q}{4pq}k^{2} + i\vec{k}\cdot(\vec{R}_{p}-\vec{R}_{q}) \!\Bigr)} 
\nonumber \\
&= c_{\alpha}^{(s)}c_{\beta}^{(p)}c_{\gamma}^{(s)}c_{\delta}^{(p)} 
M \dfrac{\pi}{2(pq)^{\!\!\tfrac{3}{2}}} \Bigl(I_{1}+I_{2}+I_{3}\Bigr)\,, %\nonumber
\label{vs:55a} % \tag{125a}
\end{align}
where $I_1$, $I_2$, $I_3$ are given by the following expressions:
\begin{align}
I_{1} &= R_{p}R_{q} \int\limits_{0}^{\infty} \Biggl[\;\;
\int\limits_{-1}^{1} e^{ik(R_{p}-R_{q})\cos{\theta_{k}}}d\cos{\theta_{k}}
\Biggr] \exp{\left( -\dfrac{p+q}{4pq} k^{2} \right)} dk \nonumber\\
&= R_{p}R_{q}\int\limits_{0}^{\infty}j_{0}(kR_{z}) \exp{\left(\! -\dfrac{p+q}{4pq} k^{2} \right)} dk 
\nonumber \\ 
&= \dfrac{R_{p}R_{q}}{R_{z}}\int\limits_{0}^{\infty}j_{0}(x)e^{-sx^{2}}dx 
=\dfrac{\pi}{2}\dfrac{R_{p}R_{q}}{R_{z}} \,\mbox{erf}\left[\dfrac{1}{2\sqrt{s}}\right]\,, 
\label{vs:56}
\end{align}
with $R_{z}=|R_{p}-R_{q}|$ and $s=\dfrac{p+q}{4pqR_{z}^{2}}$;
\begin{align}
I_{2} &= \dfrac{1}{2}i \left( \dfrac{R_{p}}{p}-\dfrac{R_{q}}{q} \right) \int\limits_{0}^{\infty} \Biggl[\;\;
\int\limits_{-1}^{1} d\cos{\theta_{k}}\ \cos{\theta_{k}}\, e^{ ik(R_{p}-R_{q})\cos{\theta_{k}} }
\Biggr] k \nonumber \\
& \qquad\qquad \times \exp{ \left( -\dfrac{p+q}{4pq}k^{2} \right) } dk  \nonumber\\
&= \dfrac{R_{q}q-R_{p}p}{pqR_{z}^{2}}\int\limits_{0}^{\infty} j_{1}(x)xe^{-sx^{2}}dx \nonumber \\
&= \dfrac{R_{q}q-R_{p}p}{pqR_{z}^{2}} \left\{
\sqrt{\dfrac{\pi}{4s}} \exp{\left(-\dfrac{1}{4s}\right)} -\dfrac{\pi}{2}\,\mbox{erf}\left(\dfrac{1}{2\sqrt{s}}\right) 
\right\}\,;
\label{vs:57}
\end{align}
\begin{align}
I_{3}&=\dfrac{1}{16pq} \int\limits_{0}^{\infty} \Biggl[\;\;
\int\limits_{-1}^{1}d\cos{\theta_{k}}\ \cos^{2}\!{\theta_{k}}\, e^{ik(R_{p}-R_{q})\cos{\theta_{k}}}
\Biggr] k^{2} 
\nonumber \\
& \qquad \qquad \times \exp{ \left( -\dfrac{p+q}{4pq}k^{2} \right) } dk 
\nonumber\\
&=\dfrac{1}{8pqR_{z}^{3}}\int\limits_{0}^{\infty} \left[  -j_{2}(x)-\dfrac{j_{1}(x)}{x}	\right] e^{-sx^{2}}x^{2} dx 
\nonumber \\
&= \dfrac{\sqrt{\pi}}{8pqR_{z}^{3}}\left[ \dfrac{1+4s}{4s^{\tfrac{3}{2}}} \exp{\left(-\dfrac{1}{4s} \right)} -\sqrt{\pi}\ \mbox{erf}\left(\dfrac{1}{2\sqrt{s}}\right) \right]\,.
\label{vs:58}
\end{align}
Thus, $D_{ab}^{(spE)}$ is given by \eqref{vs:55a} where $M$ is given by \eqref{vs:49},
$p=\alpha+\beta$, $q=\gamma+\delta$; $I_{1}$, $I_{2}$ and $I_{3}$ are given 
by Eqs~(\ref{vs:56}) -- (\ref{vs:58}),
$R_{z}=|R_{p}-R_{q}|$, $R_{p}=\dfrac{\alpha}{p}d$, $R_{q}=\dfrac{\gamma}{q}d$ and $s=\dfrac{p+q}{4pqR_{z}^{2}}$.

\noindent
Finally, the evaluation of the \textbf{Coulomb-type} integrals between $s$ and $p$
functions at different sites proceeds as follows:
\begin{align}
&D_{ab}^{(spC)}(\alpha,\beta,\gamma,\delta;d)=\int d\vec{r}_{1}d\vec{r}_{2} g_{1s}(\alpha,\vec{r_{1}}-\vec{d})g_{1s}(\beta,\vec{r_{1}}-\vec{d}) \nonumber \\ 
& \qquad \times \dfrac{1}{|\vec{r}_{1}-\vec{r}_{2}|} g_{2p} (\gamma,\vec{r}_{2})g_{2p}(\delta,\vec{r}_{2}) 
\label{vs:59} \\
& \qquad = \int \vec{r}_{1}d\vec{r}_{2} f_{1}(\vec{r}_{1}) f_{2}(r_{12}) f_{3}(\vec{r}_{2})\,, \nonumber
\end{align}
where
\begin{align*}
f_{1}(\vec{r}_{1}) &= g_{1s}(\alpha,\vec{r}_{1}-\vec{d})g_{1s}(\beta,\vec{r}_{1}-\vec{d})\,, \\
f_{2}(r_{12}) &= \dfrac{1}{|\vec{r}_{1}-\vec{r}_{2}|}\,, \\
f_{3}(\vec{r}_{2}) &= g_{2p}(\gamma,\vec{r}_{2})g_{2p}(\delta,\vec{r}_{2})\,.
\end{align*}

\noindent
The Fourier transforms of these functions are:
\begin{align}
&f_{1}(\vec{k}) = \exp{ \left( -\dfrac{\alpha\beta}{\alpha+\beta}d^{2} \right) } 
\left( \dfrac{\pi}{p} \right)^{\!\!\tfrac{3}{2}} \exp{ \left( -\dfrac{k^{2}}{4p} - i\vec{k}\cdot\vec{d} \right) }\,,
\nonumber \\
&f_{2}(\vec{k})\ \text{as in (50)}\,, \nonumber\\
&f_{3}(\vec{k}) = \int e^{i\vec{k}\cdot\vec{r}} r^{2}\cos^{2}{\theta} e^{-(\gamma+\delta)r^{2}}d\vec{r}   \nonumber \\
&=\left( \dfrac{k^{2}}{2q} \right)^{2} \dfrac{1}{k^{2}} 
\left[ \dfrac{2q}{k^{2}} - \cos^{2}{\theta_{k}} \right] \left( \dfrac{\pi}{q} \right)^{\!\!\tfrac{3}{2}} 
\exp{ \left(- \dfrac{k^{2}}{4q} \right) }\,, 
\label{vs:60}
\end{align}
as it has been calculated before, Eq.~\eqref{vs:35}. Therefore:
\begin{align*}
&D_{ab}^{(spC)} = c_{\alpha}^{(s)}c_{\beta}^{(s)}c_{\gamma}^{(p)}c_{\delta}^{(p)}
\exp{\left(-\dfrac{\alpha\beta}{\alpha+\beta}d^{2}\right)} \int 
\dfrac{d\vec{k}}{(2\pi)^{3}} e^{i(\vec{k}\cdot\vec{d})} \left( \dfrac{\pi}{p} \right)^{\!\!\tfrac{3}{2}} 
\\
&\quad\times \exp{ \left( -\dfrac{k^2}{4p} \right) } \times 
\dfrac{4\pi}{k^2}\left(\dfrac{\pi}{q} \right)^{\!\!\tfrac{3}{2}} 
\exp{ \left(\! -\dfrac{k^{2}}{4q} \!\right) } \dfrac{1}{2q} \left[1-\dfrac{k^{2}}{2q}\cos^{2}{\theta_{k}} \right] 
\\
&\quad\quad\ \ = \frac{\pi^{3}}{(pq)^{\tfrac{3}{2}}} \dfrac{2\pi}{q} \dfrac{1}{(2\pi)^{3}}
\exp{\left(\!-\dfrac{\alpha\beta}{\alpha+\beta}d^{2}\!\right)}\bigl[ X_{1}+X_{2} \bigr]\,,
\end{align*}
with
\begin{align*}
X_{1} &= \int \frac{d\vec{k}}{k^{2}} \exp{ \left( i(\vec{k}\cdot\vec{d}) -\dfrac{k^{2}}{4p} 
	- \dfrac{k^{2}}{4q} \right)} = \dfrac{2\pi^{2}}{d}\,\mbox{erf}\left( \sqrt{\dfrac{pq}{p+q}}d \right)\,, 
\\
X_{2} &= -\int \dfrac{d\vec{k}}{k^{2}} e^{i(\vec{k}\cdot\vec{d})} \frac{k^{2}}{2q}\cos^{2}{\theta_{k}} 
\exp{ \left( -\dfrac{p+q}{4pq}k^{2} \right)} 
\\
& = \dfrac{2\pi}{qd^{3}} \Biggl[ \sqrt{\pi} \dfrac{1+4s}{4s^{\tfrac{3}{2}}} 
\exp{\left(\! -\frac{1}{4s}\!\right)} -\pi\, \mbox{erf}\left(\dfrac{1}{2\sqrt{s}}\right)\Biggr]\,, 
\end{align*}
where $\; s = \dfrac{p+q}{4pqd^{2}}. \;$ Finally, we obtain:
\begin{align}
D_{ab}^{(spC)} &= c_{\alpha}^{(s)}c_{\beta}^{(s)}c_{\gamma}^{(p)}c_{\delta}^{(p)} \exp{\left(-\dfrac{\alpha\beta}{\alpha+\beta}d^{2}\right)} \nonumber \\
&\times \Bigg[
\dfrac{\pi^{3}}{2p^{\tfrac{3}{2}}q^{\tfrac{5}{2}}d}\left(1-\dfrac{1}{qd^{2}}\right)\mbox{erf}\left(
\sqrt{\frac{pq}{p+q}}d^{2} \right) \nonumber \\
& \qquad
+ \dfrac{\pi^{\tfrac{5}{2}}}{(p+q)^{\tfrac{3}{2}}q^{2}}\left(1+\frac{p+q}{pqd^{2}}\right) 
\exp{ \left(\! -\dfrac{pq}{p+q}d^{2} \!\right) } \Bigg]\,.
\label{vs:61}
\end{align}
The equations according to which the matrix elements summarized in Table~\ref{t8}
are calculated are reminded in Table~\ref{t9}.

%\newpage
%\subsubsection*{ Summary of two-electron matrix elements.}
%Following the relations in Table \ref{t6}, the matrix elements are the ones displayed at Table \ref{t5}

\begin{table}[t!]
	\caption{\label{t9}
		Equations specifying matrix elements of Table \ref{t8}.}
	\centering
	\begin{tabular}{ccccc}
		\hline
		Matrix & && Matrix & \\*[-1pt]  elements & Eq. && elements & Eq. \\
		\hline
		$D_{aa}^{(ss)}$  & \eqref{vs:44}  & \rule[-2pt]{0pt}{13pt} & $D_{ab}^{(ssE)}$ & (\ref{vs:51}) \\
		$D_{ab}^{(ssC)}$ & \eqref{vs:52a} & \rule[-2pt]{0pt}{13pt} & $D_{ab}^{(spE)}$ & (\ref{vs:55a}) -- (\ref{vs:58}) \\
		$D_{ab}^{(spC)}$ & \eqref{vs:61}  & \rule[-2pt]{0pt}{13pt} & $D_{bb}^{(spE)}$ & (\ref{vs:46}) \\
		$D_{bb}^{(spC)}$ & \eqref{vs:47}  & \rule[-5pt]{0pt}{16pt} & $D_{bb}^{(pp)}$  & (\ref{vs:45}) \\
		\hline
	\end{tabular}
\end{table}

%\begin{table}[h!]
% \centering
% \renewcommand{\arraystretch}{1.5}
% \begin{tabular}{c|c}
%  $D_{aa}^{(ss)}$ & $Eq. \ref{vs:44}$ \\ \hline
%  $D_{ab}^{(ssE)}$ & $Eq. \ref{vs:51}$ \\ \hline
%  $D_{ab}^{(ssC)}$ & $Eq. \ref{vs:52a}$ \\ \hline
%  $D_{ab}^{(spE)}$ & $Eq. \ref{vs:55a}-Eq. \ref{vs:58}$ \\ \hline
%  $D_{ab}^{(spC)}$ & $Eq. \ref{vs:61}$ \\ \hline
%  $D_{bb}^{(spE)}$ & $Eq. \ref{vs:46}$ \\ \hline
%  $D_{bb}^{(spC)}$ & $Eq. \ref{vs:47}$ \\ \hline
%  $D_{bb}^{(pp)}$ & $Eq. \ref{vs:45}$ \\ \hline
% \end{tabular}
% \caption{Relations between matrix elements of Table \ref{t5} and equations}
% \label{t6}
%\end{table}
\end{document}